\documentclass[12pt]{article}
\usepackage{graphicx}
\newcommand{\bfb}{\cal}
\newcommand{\Sp}{{\rm Sp}}
\newcommand{\mthb}{\bf }
\newcommand{\Ima}{\mathop {\rm Im}\nolimits}

\newcommand{\signum}{\mathop{\rm sign}\nolimits}

\newcommand{\bsp}{(j,\vec q,\omega_m)}

\begin{document}



{\large\bf Spin waves in nanosized magnetic films }

\bigskip
{\large L.V. Lutsev}

\bigskip
{\it A.F. Ioffe Physical-Technical Institute of the Russian
Academy of Sciences, St Petersburg, 194021, Russia\\ ${ }$\\
E-mail: l\_lutsev@mail.ru}



\begin{abstract}
We have studied spin excitations in nanosized magnetic films in the
Heisenberg model with magnetic dipole and exchange interactions by
the spin operator diagram technique. Dispersion relations of spin
waves in thin magnetic films (in two-dimensional magnetic monolayers
and in two-layer magnetic films) and the spin-wave resonance
spectrum in $N$-layer structures are found. For thick magnetic films
generalized Landau-Lifshitz equations are derived from first
principles. Landau-Lifshitz equations have the integral
(pseudodifferential) form, but not differential one. Spin
excitations are determined by simultaneous solution of the
Landau-Lifshitz equations and the equation for the magnetostatic
potential. For normal magnetized ferromagnetic films the spin wave
damping has been calculated in the one-loop approximation for a
diagram expansion of the Green functions at low temperature. In
thick magnetic films the magnetic dipole interaction makes a major
contribution to the relaxation of long-wavelength spin waves. Thin
films have a region of low relaxation of long-wavelength spin waves.
In thin magnetic films four-spin-wave processes take place and the
exchange interaction makes a major contribution to the damping. It
is found that the damping of spin waves propagating in magnetic
monolayer is proportional to the quadratic dependence on the
temperature and is very low for spin waves with small wavevectors.
Spin-wave devices on the base of nanosized magnetic films are
proposed -- tunable narrow-band spin-wave filters with high quality
at the microwave frequency range and field-effect transistor (FET)
structures contained nanosized magnetic films under the gate
electrode. Spin-wave resonances in nanosized magnetic films can be
used to construct FET structures operating in Gigahertz and
Terahertz frequency bands.
\end{abstract}

75.10.Jm; 75.30.Ds

\bigskip

Heisenberg model, diagram technique, spin waves, nanosized magnetic
films, relaxation, spin waves devices in Gigahertz and Terahertz
frequency bands

\section{Introduction}

Nanosized magnetic films are of great interest due to their
perspective applications in spin-wave devices. At present, the most
important spin waves -- microwave filters, delay lines,
signal-to-noise enhancers, and optical signal processors have been
realized on the base of magnetic films of microwave
thickness~\cite{Stan93,Stan09,Kabos}. Nanosized films give us
opportunity to construct spin-wave devices of small sizes and to
design devices with new functional properties. Recently new
applications of spin waves have been proposed -- spin-wave computing
\cite{Khitun10,Lenk11}, spin-wave filtering using width-modulated
nanostrip waveguides \cite{Kim10}, and transmission of electrical
signals by spin-wave interconversion in an insulator garnet Y${
}_3$Fe${ }_5$O${ }_{12}$ (YIG) film based on the spin-Hall effect
\cite{Kaj10}. Spin-wave logic elements have been done on the base of
a Mach-Zehnder-type interferometer \cite{Kim10,Schneider08,Liu11}
and can be realized on magnonic crystals~\cite{Lenk11}. Using
nanosized magnetic films, we have probability to construct array of
logic elements of small sizes.

In order to design new spin-wave devices based on nanosized magnetic
films, it is necessary to determine the dispersion relations and
damping of spin excitations in nanosized films. In the
phenomenological model with the magnetic dipole interaction (MDI)
and the exchange interaction \cite{Kal86,Kal90,Linear,Gur96} the
magnetization dynamics in thick magnetic films is described by the
Landau-Lifshitz equations, which are differential with respect to
spatial variables. The differential form of equations is postulated.
In this connection, the following question arises: is this form of
Landau-Lifshitz equations correct for nanosized films? Determination
of the dispersion relations depends on the answer of this question.
In phenomenological models the spin-wave damping is described by
relaxation terms in Gilbert, Landau-Lifshitz, or Bloch
forms~\cite{Gur96}. Properties of intrinsic relaxation processes are
not taken into account in these terms and, therefore, the calculated
spin-wave damping may be incorrect. The above-mentioned leads us to
the main question of the paper: what are the dispersion relations
and damping of spin waves in nanosized films and can they be derived
from first principles? In order to answer this question, we develop
the Heisenberg model with the MDI and the exchange interaction. In
the framework of this model we consider spin excitations in
nanosized films, relaxation of spin waves, and generalize
Landau-Lifshitz equations.

The above-mentioned problems have not yet been investigated
comprehensively. One of the cause of these problems is the
long-range action of the MDI. The spin-wave relaxation and the
spin-wave dynamics become dependent on the dimensions and shapes of
ferromagnetic samples. In order to analyze the Heisenberg model with
the MDI and the exchange interaction we use the spin operator
diagram technique~\cite{Izyum,Vaks67a,Vaks67b,Lut05,Lut08}.
Advantages of the spin operator diagram technique are: the
opportunity to calculate the spin wave damping at high temperatures
and more exact relationships describing spin-wave scattering and
excitations in comparison with methods based on diagram techniques
for creation and annihilation magnon Bose
operators~\cite{Erick91a,Erick91b,Mills92,Costa98,Costa2000,Per04,Kreisel09,Ng11,Mel11}.
In \cite{Lut08,Lut07} the spin operator diagram technique is
generalized for models with arbitrary internal Lie-group dynamics.

In section 2 we consider spin operator diagram technique for the
Heisenberg model with the MDI and the exchange interaction.
Spin-wave excitations are determined by poles of the ${\bfb
P}$-matrix -- the matrix of the effective Green functions and
interaction lines. On the base of this diagram technique dispersion
relations of spin waves in a normal magnetized monolayer and in a
magnetized structure consisted of two monolayers and the spectrum of
spin-wave resonances in a $N$-layer structure are found (section 3).
For thick magnetic films it is more convenient to present the ${\bfb
P}$-matrix-pole equation describing spin-wave excitations in the
form of the Landau-Lifshitz equations and the equation for the
magnetostatic potential (section 4). Spin excitations are determined
by simultaneous solution of these equations. Landau-Lifshitz
equations are integral (pseudodifferential) equations, but not
differential ones with respect to spatial variables. The reduction
of Landau-Lifshitz equations to differential equations with exchange
boundary conditions is incorrect and their solutions give dispersion
relations differed from dispersion relations calculated on the base
of integral (pseudodifferential) Landau-Lifshitz equations. In
section 5 we consider spin-wave relaxation in thick and thin
magnetic films. In thick films three-spin-wave processes take place
and the MDI makes a major contribution to the relaxation of
long-wavelength spin waves. Thin films have a region of low
relaxation of long-wavelength spin waves. In this case,
three-spin-wave processes are forbidden and the exchange interaction
makes a major contribution to the relaxation process. Nanosized
magnetic films with low relaxation spin waves are applicable to
microwave spin wave devices. Tunable narrow-band spin-wave filters
with high quality at the microwave frequency range and field-effect
transistor (FET) structures contained nanosized magnetic films under
the gate electrode are proposed in section 6. Spin-wave resonances
of nanosized magnetic films can be used to construct FET structures
operating in Gigahertz and Terahertz frequency bands.

\section{Heisenberg model with magnetic dipole and exchange interactions}
\subsection{Spin operator diagram technique}

Let us consider the Heisenberg model with the exchange interaction
and the MDI on a crystal lattice~\cite{Lut05,Lut08}. The exchange
interaction is short-ranged and the MDI is long-ranged. Operators
$S^{\pm}=S^x\pm iS^y$, $S^z$ satisfy the commutation relation

$$[S^z(\vec 1),S^{+}({\vec 1}')]=S^{+}(\vec 1)\delta_{\vec 1{\vec
1}'}$$

$$[S^z(\vec 1),S^{-}({\vec 1}')]=-S^{-}(\vec 1)\delta_{\vec 1{\vec
1}'}$$

$$[S^{+}(\vec 1),S^{-}({\vec 1}')]=2S^z(\vec 1)\delta_{\vec 1{\vec
1}'} ,$$

\noindent where $\vec 1 \equiv \vec{r}_1 , {\vec{1}}'\equiv
\vec{r_1}{\,}'$ is the abridged notation of crystal lattice sites.

\noindent The Hamiltonian of the Heisenberg model is

\begin{equation}
{\bfb H} =-g{\mu}_B \sum_{\vec 1}H({\vec 1})S^z({\vec 1})- g{\mu}_B
\sum_{\vec{1}}h_{\mu}(\vec{1})S^{\mu}(\vec{1})-\frac12\sum_{\vec{1},{\vec{1}}'}
J_{\mu\nu}(\vec{1}-\vec{1}')S^{\mu}(\vec{1})S^{\nu}({\vec{1}}'),
\label{eq1}
\end{equation}

\noindent where $H$ ($\vec H\parallel Oz$) is the external magnetic
field, $h_{\mu}$ is the auxiliary infinitesimal magnetic field,
$\mu=$ $-$, $+$, $z$. It is supposed that the summation in
(\ref{eq1}) and in the all following relations is performed over all
repeating indices $\mu$, $\nu$. The summation is carried out over
the crystal lattice sites $\vec 1 , {\vec 1}'$ in the volume $V$ of
the ferromagnetic sample. $g$ and ${\mu}_B$ are the Land\' e factor
and the Bohr magneton, respectively. $J_{\mu\nu}(\vec{1}-\vec{1}') =
J_{\nu\mu}(\vec{1}'-\vec{1})$ is the interaction between spins,
which is the sum of the exchange interaction $ I_{\mu\nu}$ and the
MDI

\begin{equation}
J_{\mu\nu}(\vec{1}-\vec{1}')=
I_{\mu\nu}(\vec{1}-\vec{1}')-\left.4{\pi}(g{\mu}_B)^2
\nabla_{\mu}{\Phi}(\vec{r}-\vec{r}{\,}')\nabla_{\nu}'\right\vert_{\vec{r}=
\vec{1},\vec{r}{\,}'=\vec{1}'}, \label{eq2}
\end{equation}
\noindent where $\Phi(\vec{r}-\vec{r}{\,}')$ is determined by the
equation

\begin{equation}
\Delta\Phi(\vec{r}-\vec{r}{\,}')=\delta(\vec{r}-\vec{r}{\,}'),
\label{eq3}
\end{equation}

\[\nabla_{\mu}=\{\nabla_{-},\nabla_{+},\nabla_z\}=\left\{\frac12\left (\frac{\partial}{\partial
x}+i\frac{\partial}{\partial y}\right),\frac12
\left(\frac{\partial}{\partial x}-i\frac{\partial} {\partial
y}\right),\frac{\partial}{\partial z}\right\}.\]

\noindent In the 3-dimensional space $\Phi(\vec{r}-\vec{r}{\,}')=
-1/4\pi|\vec{r}-\vec{r}{\,}'|$ and the MDI term in the Hamiltonian
(\ref{eq1}) can be written as

\[{\bfb H}^{(dip)} =\frac{(g{\mu}_B)^2}{2} \sum_{\vec 1,\vec{1}'}
\left[\frac{(\vec S(\vec 1),\vec
S(\vec{1}'))}{|\vec{1}-\vec{1}'|^3}- \frac{3(\vec S(\vec
1),\vec{1}-\vec{1}')(\vec
S(\vec{1}'),\vec{1}-\vec{1}')}{|\vec{1}-\vec{1}'|^5}\right].\]

\noindent For the following calculations of spin-wave dispersion
relations in magnetic films we use more convenient form of the MDI
determined by relations (\ref{eq2}), (\ref{eq3}).

Spin excitations, interaction of spin waves, spin wave relaxation
and other parameters of excitations in the canonical spin ensemble
are determined by the generating
functional~\cite{Izyum,Vas97,Lut07,Lut08}

\[ Z[h] = \Sp\exp[-\beta{\bfb H}(h)]\]
\begin{equation}
= \sum_{n=0}^{\infty}\sum_{\vec 1,\ldots,\vec n\atop
{\mu}_1,\ldots,{\mu}_n}
\int\limits_0^{\beta}\!\!\cdots\!\!\int\limits_0^{\beta}
Q^{{\mu}_1,\ldots,{\mu}_n}(\vec 1,\ldots,\vec
n,\tau_1,\ldots,\tau_n) h_{{\mu}_1}(\vec 1,\tau_1)\ldots
h_{{\mu}_n}(\vec n,\tau_n)\,d\tau_1\ldots d\tau_n ,\label{eq4}
\end{equation}

\noindent where $\beta =1/kT$, $k$ is the Boltzmann constant, $T$ is
the temperature, $h=\{h_{\mu_i}\}$. Coefficients
$Q^{{\mu}_1,\ldots,{\mu}_n}$ are proportional to the temperature
Green function without vacuum loops

$$G^{{\mu}_1\ldots {\mu}_n}(\vec 1,\ldots,\vec
n,\tau_1,\ldots,\tau_n)\equiv \langle\langle{\mthb{T}}\hat
S^{{\mu}_1}(\vec 1,\tau_1)\ldots\hat S^{{\mu}_n}(\vec
n,\tau_n)\rangle\rangle$$
\begin{equation}
= (\beta g{\mu}_B)^{-n}Z^{-1}\left.\frac{\delta^n Z[h]}{\delta
h_{{\mu}_1}(\vec 1,\tau_1)\ldots\delta h_{{\mu}_n}(\vec
n,\tau_n)}\right|_{h\to 0} , \label{eq5}
\end{equation}

\noindent where $\hat S^{\alpha}(\vec n,\tau) = \exp(\tau{\bfb
H})S^{\alpha}(\vec n)\exp(-\tau{\bfb H})$ are the spin operators in
the Euclidean Heisenberg representation, $\tau \in [0,\beta]$. ${\bf
T}$ is the $\tau$-time ordering operator. Variable $\tau$ is added
in the auxiliary field $h_{\mu}$ in order to take into account ${\bf
T}$-ordering. $\left\langle\left\langle
\ldots\right\rangle\right\rangle$ denotes averaging of spin
operators calculated with $\exp(-\beta{\bfb H}) /\Sp\exp(-\beta{\bfb
H})$. The symbol $\Sp$ denotes the trace.

The frequency representation of the expansion (\ref{eq4}) is more
convenient for calculations. The Fourier transforms of
$Q^{{\mu}_1,\ldots,{\mu}_n}$ are defined in terms of the Matsubara
frequencies $\omega^{(1)}_{m_1}=2\pi m_1/\hbar\beta$, $\ldots$,
$\omega^{(n)}_{m_n}=2\pi m_n/\hbar\beta$ \cite{Mats55} ($m_1$,
$\ldots$, $m_n$ are integers)

\[Q^{{\mu}_1,\ldots,{\mu}_n}(\vec 1,\ldots,\vec
n,\omega^{(1)}_{m_1},\ldots,\omega^{(n)}_{m_n}) \]

\begin{equation}
= \int\limits_0^{\beta}\!\!\cdots\!\!\int\limits_0^{\beta}
Q^{{\mu}_1,\ldots,{\mu}_n}(\vec 1,\ldots,\vec
n,\tau_1,\ldots,\tau_n)\exp[-i\hbar(\omega^{(1)}_{m_1}\tau_1+\ldots
+\omega^{(n)}_{m_n}\tau_n)]\,d\tau_1\ldots d\tau_n .\label{eq6}
\end{equation}

\noindent The coefficients $Q^{{\mu}_1,\ldots,{\mu}_n}$  can be
expanded with respect to the interaction
$J_{\mu\nu}(\vec{1}-\vec{1}')$ (\ref{eq2})
~\cite{Lut05,Lut08,Izyum,Vaks67a,Vaks67b,Lut07}. Each term of this
expansion is represented by a diagram constructed of propagators,
vertices, blocks and interaction lines.

\begin{figure*}
\begin{center}
\includegraphics*[scale=.55]{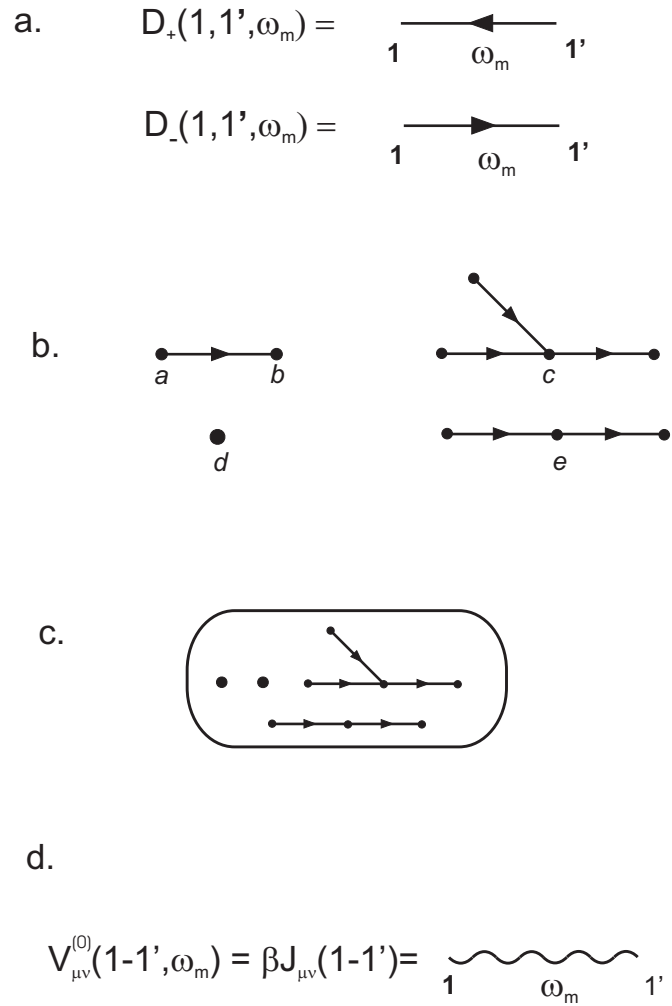}
\end{center}
\caption{(a) Propagators $D_{\pm}$, (b) vertices, (c) block with
isolated parts and (d) interaction lines  $V^{(0)}_{\mu\nu}$. }
\label{Fig1}
\end{figure*}

\noindent {\bf 1. Propagators.} Spin propagators
\begin{equation}
D_{\pm}(\vec{1},\vec{1}',\omega_m)=
\frac{\delta_{\vec{1}\vec{1}'}}{p_0\pm
i\beta\hbar\omega_m},\label{eq7}
\end{equation}

\noindent where $p_0=\beta g\mu_B H$, are determined for the spin
ensemble without any interaction between spins. The propagators
$D_{\pm}(\vec{1},\vec{1}',\omega_m)$ are represented by directed
lines in diagrams (figure~\ref{Fig1}(a)). The directions of arrows
show the direction of growth of the frequency variable $\omega_m$.

\noindent {\bf 2. Vertices.} There are five types of vertices
(figure~\ref{Fig1}(b)). Vertices $a$, $b$ are the start and end
points of propagators, respectively. In analytical expressions of
diagrams the vertex $a$ corresponds with the factor 2 and the vertex
$b$ with the factor 1. The vertex $c$ ties three propagators and
corresponds with the factor (-1) in analytical expressions. The
vertex $d$ with the factor 1 is defined as a single vertex. The
vertex $e$ ties two propagators. The factor of the $e$-vertex is
equal to (-1).

\noindent {\bf 3. Blocks.} Blocks contain propagators and isolated
vertices $d$ (figure~\ref{Fig1}(c)). Propagators can be connected
through vertices $c$, $e$. In analytical expressions of the diagram
expansion each block corresponds with the block factor
$B^{[\kappa-1]}(p_0)$, where $\kappa$ is the number of isolated
parts in the block. The factor $B^{[\kappa-1]}(p_0)$ is expressed by
partial derivatives of the Brillouin function $B_S$ for the spin $S$
with respect to $ p_0$
\begin{eqnarray}
B(p_0)&=& \langle\langle S^z\rangle\rangle_0=  SB_S(Sp_0)\nonumber\\
B^{[n]}(p_0)&=& S\frac{\partial^n B_S(Sp_0)}{\partial
p_0^n},\label{eq8}
\end{eqnarray}

\noindent where $\langle\langle\ldots\rangle\rangle_0$ denotes the
statistical averaging performed over the states described by the
Hamiltonian ${\bfb H}$ (\ref{eq1}) without the interaction
$J_{\mu\nu}$ between spins. $B_S(x)= (1+ 1/2S)\coth [(1+ 1/2S)x] -
(1/2S)\coth (x/2S)$.

\noindent {\bf 4. Interaction lines.} The interaction line
$V^{(0)}_{\mu\nu}(\vec 1-\vec{1}',\omega_m)=\beta J_{\mu\nu}(\vec
1-\vec{1}')$ connects two vertices in a diagram
(figure~\ref{Fig1}(d)). The correspondence between the first index
$\mu$ of the interaction line $V^{(0)}_{\mu\nu}$ and the vertex type
is the following. (1) If $\mu= -$, then the left end point of
$V^{(0)}_{-\nu}$ is bound to the vertex $a$; (2) if $\mu= +$, then
this end point is bound to the vertices $b$ or $c$; (3) if $\mu= z$,
then the end is bound to the vertices  $d$ or $e$. The analogous
correspondence is satisfied for the right end $\nu$ of
$V^{(0)}_{\mu\nu}$.

Coefficients $Q^{{\mu}_1,\ldots,{\mu}_n}$ in the expansion
(\ref{eq4}) in the frequency representation (\ref{eq6}) are the sum
of $N$ topologically nontrivial diagrams
$\sum_NQ^{{\mu}_1,\ldots,{\mu}_n}_N$. The general form of the
analytical expression of the diagram in the frequency representation
is written as~\cite{Lut05,Lut08,Izyum,Vaks67a,Vaks67b}
\[ Q^{{\mu}_1,\ldots,{\mu}_n}_N({\vec 1},\ldots,{\vec n}, \omega^{(1)}_{m_1}, \ldots, \omega^{(n)}_{m_n})
= (-1)^L2^{m_a}\frac{P_k}{2^kk!}\prod_lB^{[\kappa_l-1]}(p_0)
\prod^{\kappa_l}_{\vec i,\vec j\in l}\delta_{\vec i \vec j}\]

\begin{equation}
\times\sum_{{\vec 1,\ldots\vec
k\atop\vec{1}'\ldots\vec{k}'}}\sum_{m_i} V^{(0)}_{\alpha\gamma}(\vec
1-\vec{1}',\omega_{m_1})\times \ldots \times
V^{(0)}_{\rho\sigma}(\vec k-\vec{k}',\omega_{m_k})\prod_{\vec s,\vec
s{\,}'}^{I_D}D_{-}(\vec s,{\vec s}{\,}',\omega_{m_s})
\prod^{I_v}_v\delta\left(\sum_{r\in v}
\beta\hbar\omega_{m_r}\right), \label{eq9}
\end{equation}

\noindent where ${\vec 1},\ldots,{\vec n},
\omega^{(1)}_{m_1},\ldots, \omega^{(n)}_{m_n}$ are the external
lattice and frequency variables corresponded to the auxiliary fields
$h_{\mu_i}$ in the expansion (\ref{eq4}). $m_a$ is the number of
$a$-vertices in a diagram. $L$ is the number of $c$ and
$e$-vertices. $P_k$ is the number of topological equivalent
diagrams. $2k$ is the number of vertices connected with $k$
interaction lines $V^{(0)}_{\alpha\gamma}\ldots
V^{(0)}_{\rho\sigma}$. The product $\prod_l$ is performed over all
blocks of a diagram. $\kappa_l$ is the number of isolated parts in
block $l$. The term $\prod^{\kappa_l}_{\vec i,\vec j\in
l}\delta_{\vec i \vec j}$ denotes that all isolated parts in block
$l$ are determined on a single crystal lattice site. $I_D$ is the
number of propagators in a diagram. $I_v$ is the number of vertices
in a diagram. $\sum_{m_i}$ denotes the summation performed over all
inner frequency variables. The term $\prod_v^{I_v}\delta\left(
\sum_{r\in v}\beta\hbar\omega_{m_r}\right)$ gives the frequency
conservation in each vertex $v$, i.e. the sum of frequencies of
propagators and interaction lines, which come in and go out from the
vertex $v$, is equal to 0. The vertex $d$ can be connected with the
single interaction line. In the analytical expression this
corresponds to the factor $\delta(\beta\hbar\omega_m)$. The lattice
variables $\vec s,{\vec s}{\,}'$ of propagators $D_{-}$ can be inner
or external. In the first case, end points of propagators are
connected with the end points $\{\vec 1,\vec{1}',\ldots,\vec
k,\vec{k}'\}$ of interaction lines $V^{(0)}_{\alpha\gamma}\ldots
V^{(0)}_{\rho\sigma}$ and the summation $\sum_{{\vec 1,\ldots\vec
k\atop\vec{1}'\ldots\vec{k}'}}\sum_{m_i}$ is performed. In the
second case, end points of propagators are not connected with
interaction lines.

The first approximation of the diagram expansion (\ref{eq4}) is the
self-consistent field approximation, in which the effective field
acting on spins is derived and the self-consistent field
$H^{(c)}_{\mu}$ induced by the neighboring spins is taken into
account ~\cite{Izyum,Lut05,Lut08}. This leads to the substitution
$p_0\to p=\beta g\mu_B H^{(c)}_z$ in the propagator $D_{-}$ in
relation (\ref{eq7}). The self-consistent field is the sum of
exchange and magnetic dipole self-consistent fields,
$H^{(c)}_{\mu}=H^{(exch)}_{\mu}+H^{(m)}_{\mu}$, where

\begin{eqnarray}
H^{(exch)}_{\mu}(\vec 1)&=&(g\mu_B)^{-1}\sum_{\vec{1}'}I_{\mu\nu}(\vec 1-
\vec{1}')\langle\langle S^{\nu}(\vec{1}')\rangle\rangle\nonumber\\
H^{(m)}_{\mu}(\vec 1)&=&-4\pi g\mu_B\nabla_{\mu}\left.
\sum_{\vec{1}'}\Phi(\vec r-{\vec r}{\,}')
{\nabla}'_{\nu}\langle\langle
S^{\nu}(\vec{r}{\,}')\rangle\rangle\right\vert_{\vec r = \vec
1\atop\vec{r}{\,}'=\vec{1}'}.\label{eq10}
\end{eqnarray}

The second approximation of the expansion (\ref{eq4}) is the
approximation of the effective Green functions and interactions. In
this approximation, the poles of the matrix of the effective Green
functions and interactions are determined and the dispersion curves
are obtained. The next terms in the diagram expansion determine the
imaginary and real corrections to the poles of the matrix of the
effective Green functions and interactions. The imaginary parts of
the poles give the relaxation parameters of spin excitations and the
real parts determine the corrections to the dispersion curves. In
the next section we consider the approximation of the effective
Green functions and interactions.

\subsection{Effective Green functions and interaction lines}

In the framework of this approximation the matrix of the effective
Green functions and effective interactions ${\bfb P}=\Vert
P_{AB}(\vec{1},\vec{1}',\omega_m)\Vert$ is
introduced~\cite{Lut05,Lut08}. We compose the ${\bfb P}$-matrix from
analytical expressions of connected diagrams with two external
sites. These sites are end points of propagators, single vertices
$d$, or end points of interaction lines. Accordingly, multiindices
$A=(a\mu)$, $B=(b\nu)$ are the double indices, where
$\mu,\nu=\{{-},{+},z\}$ and indices $a$, $b$ point out that $A$, $B$
belong to a propagator or to a $d$-vertex $(a,b=1)$, or belong to an
interaction line $(a,b=2)$. The zero-order approximation ${\bfb
P}^{(0)}$ of the $\bfb P$-matrix is determined by the matrix of the
bare interaction ${\bfb V}^{(0)}= \Vert
V^{(0)}_{\mu\nu}(\vec{1}-\vec{1}',\omega_m)\Vert $ and by the
two-site Green functions (\ref{eq5}) in the self-consistent-field
approximation ${\bfb G}^{(0)}= \Vert G^{(0)}_{\mu\nu}\Vert $, given
on a crystal lattice site

\[{\bfb P}^{(0)} =
\left(\begin{array}{ccc} \Vert
P^{(0)}_{(1\mu)(1\nu)}\Vert&\vdots&\Vert
P^{(0)}_{(1\mu)(2\nu)}\Vert\\ \cdots&\cdots&\cdots\\ \Vert
P^{(0)}_{(2\mu)(1\nu)}\Vert&\vdots&\Vert P^{(0)}_{(2\mu)(2\nu)}\Vert
\end{array}\right)=\left(\begin{array}{ccc} \Vert G^{(0)}_{\mu\nu}\Vert&\vdots&0\\
\cdots&\cdots&\cdots\\0&\vdots&\Vert V^{(0)}_{\mu\nu}\Vert
\end{array}\right),\]
\noindent where

\begin{equation}
\Vert G^{(0)}_{\mu\nu}\Vert  = \left(\begin{array}{ccc} 0&
G^{(0)}_{{-}{+}}&0\\ G^{(0)}_{{+}{-}}&0&0\\ 0&0&\ G^{(0)}_{zz}
\end{array}\right)=\left(\begin{array}{ccc} 0& 2B(p)D_{-}(\vec 1,\vec 1',\omega_m)&0\\
2B(p)D_{+}(\vec 1,\vec 1',\omega_m)&0&0\\ 0&0&\
B^{[1]}(p)\delta_{\vec 1\vec 1'}\delta_{m0}
\end{array}\right) \label{eq11}
\end{equation}

\noindent with the propagator (\ref{eq7}), in which the substitution
$p_0\to p=\beta g\mu_B H^{(c)}_z$ is performed.

\begin{figure*}
\begin{center}
\includegraphics*[scale=0.7]{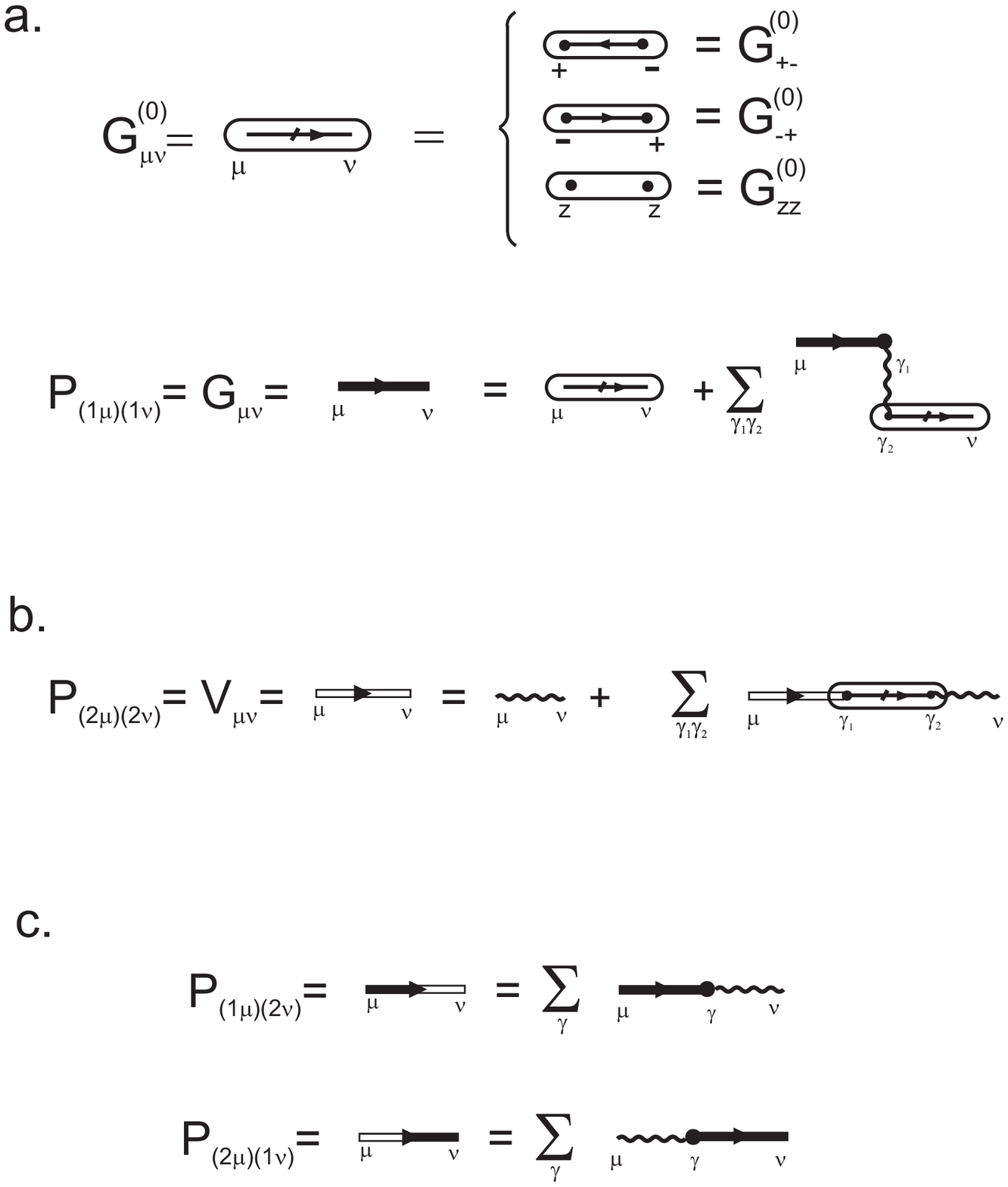}
\end{center}
\caption{(a) Definition of the effective Green functions
$P_{(1\mu)(1\nu)}=G_{\mu\nu}$ via the bare two-site Green functions
$G^{(0)}_{\mu\nu }$. (b) Definition of effective interaction lines
$P_{(2\mu)(2\nu)}= V_{\mu\nu}$. (c) Definition of intersecting terms
$P_{(1\mu)(2\nu)}$, $P_{(2\mu)(1\nu)}$.} \label{Fig2}
\end{figure*}

The $\bfb P$-matrix is obtained by means of the summation of the
${\bfb P}^{(0)}$-matrix -- the summation of all diagram chains
consisted of the bare Green functions $G^{(0)}_{\mu\nu}$ and the
bare interaction lines $V^{(0)}_{\mu\nu}$ (figure~\ref{Fig2}). These
chains of propagators and interaction lines do not have any loop
insertion. Analytical expressions of the considered  diagrams can be
written in accordance with relation~(\ref{eq9}). The summation gives
equation of the Dyson type, which forms the relationship between
${\bfb P}^{(0)}$- and ${\bfb P}$-matrices
\begin{equation}
{\bfb P}= {\bfb P}^{(0)}+ {\bfb P}\sigma{\bfb P}^{(0)}, \label{eq12}
\end{equation}
\noindent where
\[\sigma=\left(\begin{array}{ccc}
0&\vdots&{\bfb E}\\ \cdots&\cdots&\cdots\\ {\bfb E}&\vdots&0
\end{array}\right),\]
\noindent ${\bfb E}=\Vert\delta_{\mu\nu}\Vert$ is the diagonal
matrix.

The $\bfb P$-matrix consists of the two-site effective Green
functions ${\bfb G}= \Vert G_{\mu\nu}\Vert $ $=$ ${\bfb
G}^{(0)}({\bfb E}- {\bfb V}^{(0)} {\bfb G}^{(0)})^{-1}$, where
$G_{\mu\nu}= P_{(1\mu)(1\nu)}$, effective interactions ${\bfb V}=
\Vert V_{\mu\nu}\Vert $ $=$ ${\bfb V}^{(0)} ({\bfb E}- {\bfb
G}^{(0)}{\bfb V}^{(0)})^{-1}$, where $V_{\mu\nu}= P_{(2\mu)(2\nu)}$,
and intersecting terms  $P_{(1\mu)(2\nu)}$, $P_{(2\mu)(1\nu)}$
(figure~\ref{Fig2}). The effective Green functions, effective
interactions and intersecting terms are denoted in diagrams by
directed thick lines, empty lines and compositions of the thick line
- empty line, respectively. The $\bfb P$-matrix determines the
spectrum of quasi-particle excitations in the spin ensemble.
Spectrum relations for spin excitations are given by the $\bfb
P$-matrix poles -- by zero eigenvalues of the operator
$1-\sigma{\bfb P}^{(0)}$ or, equivalently, by ${\bfb E}- {\bfb
V}^{(0)} {\bfb G}^{(0)}$ under the analytical continuation
\[i\omega_m\to\omega+i\varepsilon\signum\omega\]
\begin{equation}
\delta(\beta\hbar\omega_m)=
\delta_{m0}\to\frac1{\beta\hbar(\omega+i\varepsilon\signum\omega)}
\qquad(\varepsilon\to+0). \label{eq13}
\end{equation}

Since, zero eigenvalues of the operator ${\bfb E}- {\bfb V}^{(0)}
{\bfb G}^{(0)}$ can corresponds to different eigenfunctions and can
determine different excitation modes, we introduce the spectral
parameter $\lambda$ for eigenfunctions $h^{(\lambda)}_\mu(\vec
1,\omega_m)$ of the operator ${\bfb E}- {\bfb V}^{(0)} {\bfb
G}^{(0)}$. The spectral parameter $\lambda$ can be discrete or
continuous. Taking into account the above-mentioned, we get the
equation describing spin-wave excitations

\begin{equation}
h^{(\lambda)}_\mu(\vec 1,\omega_m)-\sum_{{\vec 1}',{\vec
1}'',\nu,\rho} V^{(0)}_{\mu\nu}(\vec 1-{\vec 1}',\omega_m)
\left.G^{(0)}_{\nu\rho}({\vec 1}',{\vec 1}'',\omega_m)
h^{(\lambda)}_\rho({\vec 1}'',\omega_m) \right|_{i\omega_m \to
\omega +i\varepsilon{\rm sign}\omega} =0 . \label{eq14}
\end{equation}

\section{Spin waves in nanosized magnetic films }

\subsection{Spin-wave equations for magnetic films}

Let us consider spin waves with the wavevector $\vec q$ in a normal
magnetized film consisted of $N$ monolayers at low temperature.
$x$-, $y$-axes are in the monolayer plane and the $z$-axis is normal
to monolayers. The external magnetic field $H$ is normal to
monolayers and is parallel to the $z$-axis. At low temperature
derivatives of the Brillouin function in $B^{[n]}(p)$ in relation
(\ref{eq8}) tend to 0 exponentially with decreasing temperature.
Thus, it follows that diagrams containing blocks with isolated parts
can be dropped, the Green function $G^{(0)}_{zz}$ in relation
(\ref{eq11}) is negligible and only the Green functions
$G^{(0)}_{-+}$, $G^{(0)}_{+-}$ are taken into account in equation
(\ref{eq14}). Indices $\mu$, $\nu$  of interactions
$V^{(0)}_{\mu\nu}$ in equation (\ref{eq14}) are $\{-,+\}$. We
suppose that on monolayers spins are placed on quadratic crystal
lattice sites with the lattice constant $a$ and spin orientation is
parallel to the $z$-axis. The exchange interaction acts between
neighboring spins and is isotropic between spins in monolayers,
$2I_{{-}{+}}= 2I_{{+}{-}}= I_{zz}=I_0$, and between neighboring
layers, $2I_{{-}{+}}= 2I_{{+}{-}}= I_{zz}=I_d$. Then, the Fourier
transform of the exchange interaction with respect to the
longitudinal lattice variables $\vec 1_{xy}$ is

\[\tilde{I}(\vec q,1_z-1'_z)=\sum_{\vec 1_{xy}-\vec 1'_{xy}}I(\vec 1_{xy}-\vec
1'_{xy},1_z-1'_z)\exp[-i\vec q(\vec 1_{xy}-\vec 1'_{xy})]\]

\[= I(0,1_z-1'_z)+ 2I_0[\cos(q_xa)+\cos(q_ya)]\delta_{1_z1'_z}, \]

\noindent where $\vec 1_{xy}$, $\vec 1'_{xy}$ are crystal lattice
sites in a monolayer, $1_z$, $1'_z$ are $z$-positions of layers,
$\vec q=(q_x,q_y)$ is the longitudinal wavevector in monolayers,
$I(0,1_z-1'_z)$ is the exchange interaction at $\vec q=0$, which is
equal to $I_d$ between spins of neighboring layers. The
corresponding exchange part of the interaction line
$V^{(0)}_{\mu\nu}= V^{(exch)}_{\mu\nu}+V^{(dip)}_{\mu\nu}$
(figure~\ref{Fig1}(d)) is

\begin{equation}
V^{(exch)}_{\mu\nu}(\vec q,1_z-1'_z) =\beta\tilde{I}(\vec
q,1_z-1'_z). \label{eq15}
\end{equation}

The MDI part $V^{(dip)}_{\mu\nu}$ is determined by the Fourier
transform of equation (\ref{eq3})

\[\left(-q^2+\frac{\partial^2}{\partial
z^2}\right)\Phi(\vec{q},z-z')=S_a^{-1}\delta(z-z')\]

\noindent with the solution

\begin{equation}
\Phi(\vec q,1_z-1'_z) =\left.\Phi(\vec q,z-z')\right\vert_{z=1_z,
z'=1'_z}=\frac{-1}{2qS_a}\exp(-q|1_z-1'_z|), \label{eq16}
\end{equation}

\noindent where $S_a=a^2$, $q=|\vec q|$. According to the solution
(\ref{eq16}), the corresponding MDI part of the interaction line is

\begin{equation}
V^{(dip)}_{\mu\nu}(\vec q,1_z-1'_z)
=\frac{-4\pi\beta(g\mu_B)^2q_{\mu}q_{\nu}}{qS_a}\exp(-q|1_z-1'_z|),\qquad
(\mu,\nu=\{{-},{+}\}).\label{eq17}
\end{equation}

\noindent where
\[q_{-}=\frac{1}{2}(q_x+iq_y)\qquad q_{+}=\frac{1}{2}(q_x-iq_y)\]

\noindent Taking into account relations (\ref{eq15}) and
(\ref{eq17}), from equation (\ref{eq14}) we obtain equations for
spin-wave modes with the wavevector $\vec q$ in $N$-layer magnetic
films

\[h^{(\lambda)}_\mu(\vec q,1_z,\omega_m)- \sum_{{\vec 1}'_z}
\left[V^{(0)}_{\mu{-}}(\vec q,1_z-1'_z,\omega_m)
G^{(0)}_{{-}{+}}(1'_z,1'_z,\omega_m)h^{(\lambda)}_{+}(\vec
q,1'_z,\omega_m)\right.\]

\begin{equation}
+ \left.\left.V^{(0)}_{\mu{+}}(\vec q,1_z-1'_z,\omega_m)
G^{(0)}_{{+}{-}}(1'_z,1'_z,\omega_m) h^{(\lambda)}_{-}(\vec
q,1'_z,\omega_m)\right] \right|_{i\omega_m \to \omega
+i\varepsilon{\rm sign}\omega} =0 , \label{eq18}
\end{equation}

\noindent where

\[G^{(0)}_{{{-}{+}\atop({+}{-})}}(1_z,1'_z,\omega_m)=\frac{2B(p)\delta_{1_z1'_z}}{p\pm
i\beta\hbar\omega_m},\]

\noindent $\lambda=1,\ldots,N$ is the mode number,
$V^{(0)}_{\mu\nu}(\vec q,1_z-1'_z,\omega_m)=V^{(exch)}_{\mu\nu}(\vec
q,1_z-1'_z)+ V^{(dip)}_{\mu\nu}(\vec q,1_z-1'_z)$,
$\mu,\nu=\{{-},{+}\}$. Eigenvalues of equations (\ref{eq18}) give
dispersion relations of spin waves. In next sections we find
spin-wave dispersion relations for the cases of monolayer and
two-layer films and spin-wave resonance relations for the case of
$N$-layer structures.

\subsection{Spin waves in magnetic monolayer}

Dispersion relations of spin waves in normal magnetized monolayer
are determined by the determinant of equations (\ref{eq18}) for
variables $h^{(1)}_{-}$, $h^{(1)}_{+}$. Taking into account
relations (\ref{eq15}) and (\ref{eq17}), we find

\begin{equation}
\omega^2(\vec q)=\Omega(\vec q)[\Omega(\vec q)+
2\pi\gamma\sigma_mq], \label{eq19}
\end{equation}

\noindent where

\[\Omega(\vec q)=\gamma(H+H^{(m)})+\frac{2B(p)I_0}{\hbar}[2-\cos(q_xa)-
\cos(q_ya)],\]

\begin{figure*}
\begin{center}
\includegraphics*[scale=0.4]{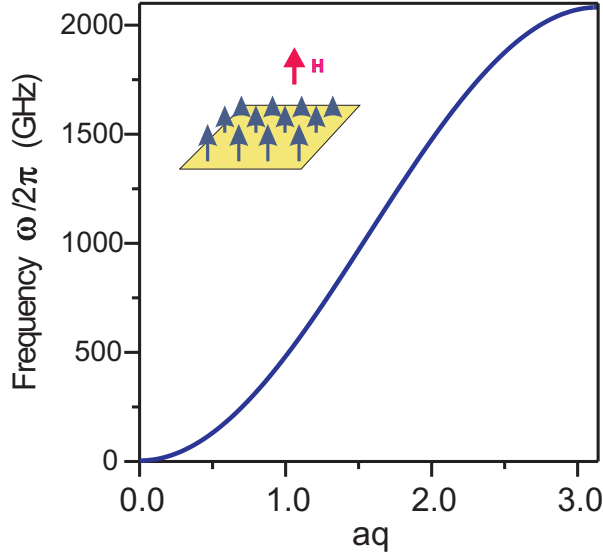}
\end{center}
\caption{Dispersion curve of spin waves propagating in the normal
magnetized monolayer film with cubic lattice ($a=$ 0.4~nm) at the
sum of magnetic fields $H+H^{(m)}=$ 3~kOe. Exchange interaction
$I_0$ is 0.085~eV. } \label{Fig3}
\end{figure*}

\noindent $\gamma=g\mu_B/\hbar$ is the gyromagnetic ratio, $H^{(m)}$
is the depolarizing magnetic field (\ref{eq10}),
$\sigma_m=g\mu_BB(p)/S_a$ is the surface magnetic moment density,
$q=(q_x^2+q_y^2)^{1/2}$. As one can see from relation (\ref{eq19}),
in monolayer films spin waves have the one-mode character. Figure
\ref{Fig3} presents the dispersion curve (\ref{eq19}) of spin waves
propagating in the monolayer film with the lattice constant $a=$
0.4~nm. The spin-wave wavevector $\vec q$ is parallel to the
$x$-axis ($q_x=q$, $q_y=0$) and is in the range $[0,\pi/a]$.
Calculations have been done for the exchange interaction between
neighboring spins $I_0=$ 0.085~eV, $B(p)=1/2$ at the sum of magnetic
fields $H+H^{(m)}=$ 3~kOe. The exchange interaction makes a major
contribution to the dispersion. The relatively weak MDI is
significant for the dispersion at small values of the wavevector
$\vec q$.

\subsection{Spin waves in two-layer magnetic film}

Let us consider spin waves in a normal magnetized structure
consisted of two monolayers of the quadratic lattice with the
lattice constant $a$. The distance between layers is equal to $d$
and the exchange interaction between spins of layers is $I_d$.
Dispersion relations are determined by eigenvalues of equations
(\ref{eq18}) for  variables $h^{(1)}_{-}$, $h^{(1)}_{+}$,
$h^{(2)}_{-}$, $h^{(2)}_{+}$ and can be written as

\[\omega^{(n)2}(\vec q)=\Omega(\vec q)[\Omega(\vec q)+
2\pi\gamma\sigma_mq]+\frac{2B(p)I_d}{\hbar}\left[\frac{2B(p)I_d}{\hbar}-
2\pi\gamma\sigma_mq\exp(-qd)\right]\]
\begin{equation}
\pm 2\left[-\frac{2B(p)I_d}{\hbar}(\Omega(\vec q)+
\pi\gamma\sigma_mq)+ \pi\gamma\sigma_mq\exp(-qd)\Omega(\vec
q)\right], \label{eq20}
\end{equation}

\noindent where

\[\Omega(\vec q)=\gamma(H+H^{(m)})+\frac{2B(p)I_0}{\hbar}[2-\cos(q_xa)-
\cos(q_ya)]+\frac{B(p)I_d}{\hbar},\]

\begin{figure*}
\begin{center}
\includegraphics*[scale=0.4]{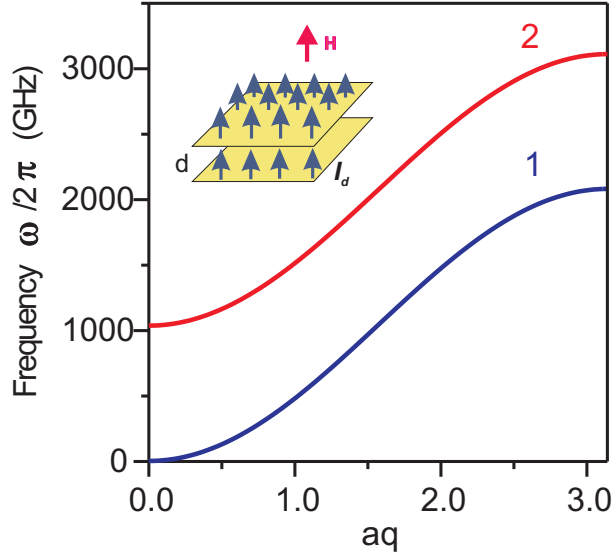}
\end{center}
\caption{Dispersion curve of spin waves propagating in the normal
magnetized two-layer magnetic  film with quadratic lattice ($a=$
0.4~nm) at the sum of magnetic fields $H+H^{(m)}=$ 3~kOe. Exchange
interactions $I_0=I_d=$ 0.085~eV. Distance between monolayers $d$ is
equal to the lattice constant $a$. 1, 2 are the first and the second
modes of spin waves, respectively. } \label{Fig4}
\end{figure*}

\noindent $n=1,2$ is the mode number, $q=(q_x^2+q_y^2)^{1/2}$. For
the first mode spins in different layers change their orientations
in-phase. In this case, spin waves of the first mode correspond to
spin waves in monolayer (\ref{eq19}). For the second mode spins in
different layers change orientations in-anti-phase and the energy of
the spin wave with the given longitudinal wavevector $q$ is higher
than the energy of the spin wave of the first mode. Dispersion
curves of spin waves determined by relations (\ref{eq20}) are shown
in figure \ref{Fig4}. Spin waves propagate along the $x$-axis.
Calculations have been done for the exchange interactions $I_0=I_d=$
0.085~eV and for the distance between layers $d=a=$ 0.4~nm at the
sum of magnetic fields $H+H^{(m)}=$ 3~kOe.

\subsection{Spin-wave resonance in $N$-layer structure}

In this section we consider spin-wave resonance in a structure
consisted of $N$ uniform monolayers with the exchange interaction
$I_d$ between spins of layers and with the distance $d$ between
layers. Spin-wave resonance is the limit case of a spin wave, when
the longitudinal wavevector $q\rightarrow 0$. Therefore, the MDI
terms $V^{(dip)}_{\mu\nu}(\vec q,1_z-1'_z)$ in equations
(\ref{eq18}) can be dropped, equations with variables
$h_{+}^{(\lambda)}$ and $h_{-}^{(\lambda)}$ are separated and
eigenvalues are determined by zero values of the determinant (we
write the determinant ${\bfb D}^{(+)}$ for equations with the
$h_{+}^{(\lambda)}$)

\[{\bfb D}^{(+)} =G^{(0)}(1)\ldots G^{(0)}(N)\]
\[\times\det\left(\begin{array}{cccc}
(G^{(0)-1}(1)-V^{(0)}(11))&-V^{(0)}(12)&0 &\vdots\\ -V^{(0)}(21)&
(G^{(0)-1}(2)-V^{(0)}(22))& -V^{(0)}(23)&
\vdots\\ 0& -V^{(0)}(32)& (G^{(0)-1}(3)-V^{(0)}(33))& \vdots\\
\cdots&\cdots&\cdots&\cdots
\end{array}\right),\]

\noindent where $V^{(0)}(ij)$, $G^{(0)}(i)$ are the abridged
notation of $V^{(exch)}_{{+}{-}}(\vec q,i_z-j_z,\omega_m)|_{\vec q
=0}$ and $G^{(0)}_{{-}{+}}(i,i,\omega_m)$ at ${i\omega_m \to \omega
+i\varepsilon{\rm sign}\omega}$, respectively. $(i,j)$ are indices
of layers. Taking into account that spins of outer layers ($i=1,N$)
interact with spins of one inner layer and spins of inner layers
interact with spins of two layers and introducing the variable for
inner layers in the determinant ${\bfb D}^{(+)}$

\[x=\frac{G^{(0)-1}(i)-V^{(0)}(ii)}{-V^{(0)}(ji)}=\frac{\hbar}{B(p)I_d}[\omega-\gamma(H+H^{(m)})]
-2\qquad (i\neq 1,N,\quad j=i\pm 1),\]

\noindent we obtain that the spin-wave resonance spectrum is
determined by roots of the polynomial

\[R_N(x)=\left|\begin{array}{ccccccc}
(x+1)&1&0&0 &\vdots &0&0\\1&x&1&0 &\vdots &0&0 \\0&1&x&1 &\vdots
&0&0 \\0&0&1&x &\vdots &0&0 \\
\cdots&\cdots&\cdots&\cdots&\cdots&\cdots&\cdots\\0&0&0&0&\vdots
&x&1 \\0&0&0&0&\vdots &1&(x+1)
\end{array}\right| \]
\[=(x+1)^2P_{N-2}(x)-2(x+1)P_{N-3}(x)+P_{N-4}(x)= 0\qquad (N\geq 2),\]

\noindent where $P_{-2}(x)=-1$, $P_{-1}(x)=0$, $P_0(x)=1$,
$P_N(x)=xP_{N-1}(x)-P_{N-2}(x)$. Polynomial $R_N(x)$ has $N$ roots

\[x^{(n)}=-2\cos\left(\frac{\pi n}{N}\right),\]

\noindent where $n=0,1,\ldots,N-1$. Taking into account the form of
the roots $x^{(n)}$, we can introduce the transverse wavevector
$q_z^{(n)}=\pi n/Nd$. Then, the spin-wave resonance spectrum can be
written as

\begin{equation}
\omega^{(n)}= \gamma(H+H^{(m)})+\frac{2B(p)I_d}{\hbar}[1-
\cos(q_z^{(n)}d)]. \label{eq21}
\end{equation}

\begin{figure*}
\begin{center}
\includegraphics*[scale=0.4]{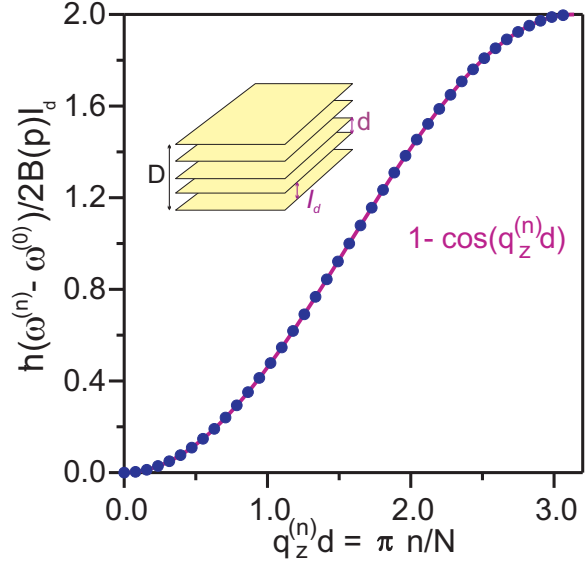}
\end{center}
\caption{Spin-wave resonance spectrum $\omega^{(n)}$
($n=0,1,\ldots,N-1$) for the structure with $N=40$ layers.
$q_z^{(n)}$ is the transverse wavevector, $d$ is the distance
between layers, $I_d$ is the exchange interaction between spins of
layers. } \label{Fig5}
\end{figure*}

\noindent For the first mode ($n=0$) spins in different layers
change their orientations in-phase. For the highest mode ($n=N-1$)
spins in different layers change orientations in-anti-phase and the
energy of spin-wave resonance is highest. Figure \ref{Fig5} presents
the spin-wave resonance spectrum for the structure with $N=40$
layers. One can see that at low values of the transverse wavevector
the resonance spectrum is proportional to the quadratic dependence
on $q_z^{(n)}$.

\section{Landau-Lifshitz equations and spin-wave excitations in
thick magnetic films}

\subsection{ Linearized Landau-Lifshitz equations}

Equations (\ref{eq14}), (\ref{eq18}) describe spin-wave excitations.
Solutions of these equations for magnetic samples of great volumes
and for thick $N$-layer magnetic films with $N\gg 1$ become
difficult, because determinants of equations (\ref{eq14}),
(\ref{eq18}) have high orders. In order to overcome the difficulty
and to find spin-wave spectrum for these samples, we derive
Landau-Lifshitz equations~\cite{Lut05,Lut08}. Dispersion relations
for spin excitations are determined by the $\bfb P$-matrix poles
(\ref{eq12}) which coincide with poles of the matrix ${\bfb G}$ of
effective propagators. Accordingly, the dispersion relations can be
derived from the eigenvalues of equation

\begin{equation} {\bfb G}= {\bfb
G}^{(0)}+{\bfb G}({\bfb V}^{({\rm exch})}+ {\bfb V}^{({\rm
dip})}){\bfb G}^{(0)}, \label{eq22}
\end{equation}

\noindent where ${\bfb G}^{(0)}= \Vert G^{(0)}_{\mu\nu}\Vert$ is the
matrix of bare propagators~(\ref{eq11}). Since the considered
interaction is the sum of exchange and magnetic dipole interactions,
we can obtain the eigenvalues and eigenfunctions of equation
(\ref{eq22}) by a two-step procedure. In the first stage, we perform
the summation of diagrams, taking into account the exchange
interaction, and find the propagator matrix ${\bfb G}^{(1)}= \Vert
G^{(1)}_{\mu\nu}\Vert $
\begin{equation}
{\bfb G}^{(1)}= {\bfb G}^{(0)}+{\bfb G}^{(0)}{\bfb V}^{({\rm
exch})}{\bfb G}^{(1)}. \label{eq23}
\end{equation}

In the second stage, the summation of diagrams with dipole
interaction lines is performed. This gives the equation for the
matrix ${\bfb G}$ of effective propagators expressed in terms of the
matrix ${\bfb G}^{(1)}$
\begin{equation}
{\bfb G}= {\bfb G}^{(1)}+{\bfb G}{\bfb V}^{({\rm dip})}{\bfb
G}^{(1)}. \label{eq24}
\end{equation}

Thus, the solution of equation~(\ref{eq22}), which determines the
matrix ${\bfb G}$, is equivalent to the solution of
equations~(\ref{eq23}), (\ref{eq24}). After the performed two-step
summation, equation (\ref{eq14}) for eigenfunctions $
h^{(\lambda)}_{\mu}$ is written in the more convenient form
\begin{equation}
h^{(\lambda)}_{\mu}(\vec{1},\omega_m)-
\left.\sum_{\rho,\sigma\atop\vec{1'}\vec{1''}} V^{({\rm
dip})}_{\mu\rho}(\vec{1}-\vec{1'}, \omega_m)
G^{(1)}_{\rho\sigma}(\vec{1'},\vec{1''}, \omega_m)
h^{(\lambda)}_{\sigma}(\vec{1''},\omega_m)\right|_{i\omega_m \to
\omega +i\varepsilon{\rm sign}\omega}= 0. \label{eq25}
\end{equation}

The solution of simultaneous equations (\ref{eq23}), (\ref{eq25})
gives the dispersion relations for spin excitations. These equations
can be reduced to linearized Landau-Lifshitz equations in the
generalized form and the equation for the magnetostatic potential.
In order to perform this transformation one needs to make a
transition to the retarded Green functions. We transform matrix
equation (\ref{eq23}) to equations describing small variations of
the magnetic moment density (or the variable magnetization),
$m_{\nu}$. The variable magnetization $m_{\nu}$ under the action of
the magnetic field $\bar h_{\nu}$, which is generated by the MDI
${\bfb V}^{({\rm dip})}$, is given by the retarded Green functions,
which are determined by the analytical continued values of the
propagator matrix ${\bfb G}^{(1)}$~\cite{Zub}
\begin{equation}
m_{\nu}(\vec{1},\omega)= \frac{\beta
(g\mu_B)^2}{V_a}\left.\sum_{\rho,\vec{1'}}
G^{(1)}_{\nu\rho}(\vec{1},\vec{1'}, \omega_m)\right |_{
i\omega_m\to\omega-i\varepsilon} \bar h_{\rho}(\vec{1'},\omega),
\label{eq26}
\end{equation}

\noindent where $V_a$ is the atomic volume. The analytical
continuation $ i\omega_m\to\omega-i\varepsilon $ defines the
retarded Green functions. $\bar h_{\rho}(\vec{1},\omega)$ is the
field of the magnetic dipole-dipole interaction acting on spins. By
multiplying matrix equation (\ref{eq23}) by ${\bfb G}^{(0)-1}$ from
the left and by $\bar h_{\rho}$ from the right, performing the
analytical continuation $ i\omega_m\to\omega-i\varepsilon $,
$\delta(\beta\hbar\omega_m) \to [\beta\hbar(\omega-
i\varepsilon)]^{-1}$ and taking into account relation (\ref{eq26}),
we get matrix equation (\ref{eq23}) in the form of simultaneous
equations
\begin{equation}
\sum_{\nu,\vec{1'}}[ G^{(0)-1}_{\rho\nu}(\vec{1},\vec{1'}, \omega) -
\beta I_{\rho\nu}(\vec{1}-\vec{1'})] m_{\nu}(\vec{1'},\omega)=
\frac{\beta (g\mu_B)^2}{V_a} \bar h_{\rho}(\vec{1},\omega).
\label{eq27}
\end{equation}

\noindent For isotropic exchange interaction, $2I_{{-}{+}}=
2I_{{+}{-}}=I_{zz}= I$, equations (\ref{eq27}) have the form
\begin{equation}
\hat E_{\pm} m_{\pm}(\vec{1},\omega)=2\gamma M(\vec{1}) \bar
h_{\mp}(\vec{1},\omega) \label{eq28}
\end{equation}

\begin{equation}
\hat E_z m_z(\vec{1},\omega)= \frac{B^{[1]}(p)}{B(p)}\gamma
M(\vec{1}) \bar h_z(\vec{1},\omega), \label{eq29}
\end{equation}

\noindent where $ M(\vec{1})= g\mu_B B(p)/V_a$ is the magnetic
moment density at the low-temperature approximation. We say that the
operators $\hat E_{\pm}$, $\hat E_z$:

\[\hat E_{\pm} m_{\pm}(\vec{1},\omega)= [\gamma(H(\vec{1})+ H^{(m)}(\vec{1}))\pm\omega] m_{\pm}(\vec{1},\omega) \]
\[+\frac{B(p)}{\hbar V_b} \sum_{\vec{1'}}\int\limits_{V_b}[\bar I(0)-\bar I(\vec q)]
\exp[i\vec q(\vec{1}-\vec{1'})] m_{\pm}(\vec{1'},\omega)\,d^3q\]

\[\hat E_z m_z(\vec{1},\omega)= \omega m_z(\vec{1},\omega)- \frac{B^{[1]}(p)}{\hbar V_b}\sum_{\vec{1'}}
\int\limits_{V_b} \bar I(\vec q) \exp[i\vec q(\vec{1}-\vec{1'})]
m_z(\vec{1'},\omega)\,d^3q\]

\noindent are Landau-Lifshitz operators. $\bar I(\vec q) =
\sum_{\vec{1}} I(\vec 1)\exp(-i\vec q\vec{1})$ is the Fourier
transform of the exchange interaction with respect to the lattice
variables. The field $ H^{(m)}(\vec{1})$ is defined by relation
(\ref{eq10}) and depends on the magnetic moment density
$M(\vec{1})$; $V_b= (2\pi)^3/V_a$ is the volume of the first
Brillouin zone. Equations (\ref{eq28}), (\ref{eq29}) have the
generalized form of the Landau-Lifshitz equations~\cite{Gur96}.
Solutions $m_{\pm}$ of equations (\ref{eq28}) depend on temperature,
because $\beta = 1/kT$ is contained in the variable $p$ of the
function $B(p)$ (\ref{eq8}), through which the magnetic moment
density $M(\vec{1})$ is expressed. Equation (\ref{eq29}) describes
longitudinal variations of the variable magnetization under the
influence of the field $\bar h_z $. At low temperature the
derivative of the function $B^{[1]}(p)$ tends to 0 and the
longitudinal variable magnetization $m_z$ is negligible.

From the form of the magnetic dipole interaction in relations
(\ref{eq2}), (\ref{eq3}) it follows that the field $\bar h_{\nu}$ in
relation (\ref{eq26}) is magnetostatic, i.e. it is expressed in
terms of the magnetostatic potential $\varphi$: $\bar h_{\nu}=
-\nabla_{\nu}\varphi $. We transform equation (\ref{eq25}) to the
equation for the magnetostatic potential $\varphi(\vec{r},\omega)$.
Taking into account relation (\ref{eq26}) and the explicit form of
the magnetic dipole interaction in relations (\ref{eq2}),
(\ref{eq3}), performing the derivation $\nabla_{\mu}$, the
analytical continuation $i\omega_m\to\omega-i\varepsilon $ and the
summation of equation (\ref{eq25}) over the index $\mu$, we obtain
the equation expressed in terms of $\varphi $, $m_{\nu}$
\begin{equation}
-\Delta\varphi(\vec{r},\omega)+
4\pi\nabla_{\nu}m_{\nu}(\vec{1},\omega)|_{\vec{1}\to\vec{r}} =0.
\label{eq30}
\end{equation}

\noindent Thus, in consideration of the Landau-Lifshitz equations
(\ref{eq28}), (\ref{eq29}), the dispersion relations of spin
excitations are given by eigenvalues of equation~(\ref{eq30}).

\subsection{ Exchange boundary conditions}

If the scale of the spatial distribution of the variable
magnetization $m_{\nu}(\vec{1},\omega)$ and the sample size are much
greater than the lattice constant $a$, then the sum over the lattice
variables $\sum_{\vec{1}}$ in Landau-Lifshitz operators $\hat
E_{\pm}$, $\hat E_z$ can be converted into an integral over the
sample volume $V_a^{-1}\int \,d^3r$. Let us consider the case when
the temperature is low and the Fourier transform of the exchange
interaction is $\bar I(\vec q) = \bar I(0) -wq^2$. Then, we obtain
that $m_z\to 0$ and equation (\ref{eq29}) is dropped. The operators
$\hat E_{\pm}$ are pseudodifferential operators of order
2~\cite{Trev}

\[\hat E_{\pm} m_{\pm}(\vec{r},\omega)= [\gamma(H(\vec{r})+ H^{(m)}(\vec{r}))\pm\omega] m_{\pm}(\vec{r},\omega) \]
\begin{equation}
+\frac{4\pi\gamma\alpha M(\vec{r})}{(2\pi)^3}
\int\limits_{V}\int\limits_{V_b}q^2 \exp[i\vec q(\vec{r}-\vec{r'})]
m_{\pm}(\vec{r'},\omega)\,d^3q\,d^3r, \label{eq31}
\end{equation}

\noindent where $\alpha=wV_a/4\pi (g\mu_B)^2$ is the exchange
interaction constant, $V$ is the volume of the ferromagnetic sample.
In \cite{Kal86,Kal90,Linear,Gur96,Rojd93,Dem01,Gus02,Gus05,Grig09}
the pseudodifferential Landau-Lifshitz operators are reduced to the
differential operators with respect to spatial variables

\begin{equation}
\hat E_{\pm}(\vec{r},\omega)= \gamma[H(\vec{r})+ H^{(m)}(\vec{r})-
4\pi\alpha M(\vec{r})\Delta] \pm\omega. \label{eq32}
\end{equation}

\begin{figure*}
\begin{center}
\includegraphics*[scale=0.4]{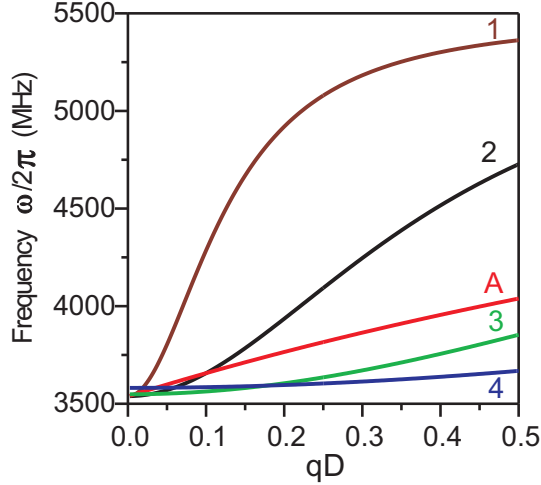}
\end{center}
\caption{ Dispersion curves of the first spin-wave mode propagating
in the YIG film of the thickness $D=$ 0.5~$\mu$m with $4\pi M=$
1750~Oe, $\alpha=$ 3.2$\cdot$10${ }^{-12}$~cm${ }^2$ at the applied
magnetic field $H=$ 3000~Oe.  The curve $A$ is calculated on the
base of relation (\ref{eq35}) for the case of pseudodifferential
Landau-Lifshitz operators (\ref{eq31}). Curves 1 - 4 are calculated
for the case of differential Landau-Lifshitz operators (\ref{eq32})
on the base of relation (\ref{eq36}) with different pinning
parameters $\xi$. (1) $\xi D$ = 0.01, (2) 0.1, (3) 1, (4) 10.}
\label{Fig6}
\end{figure*}

For solvability of equations (\ref{eq28}) with differential
Landau-Lifshitz operators (\ref{eq32}) the exchange boundary
conditions are imposed

\[\frac{\partial m_{\nu}}{\partial \vec n} +\xi m_{\nu}|_{\partial V}=0,\]

\noindent where $\vec n$ is the inward normal to the boundary
$\partial V$, and $\xi $ is the pinning parameter. This reduction to
differential Landau-Lifshitz operators is not correct. Figure
\ref{Fig6} presents exact dispersion relations of spin excitations
given by eigenvalues of equations (\ref{eq28}), (\ref{eq30}) with
pseudodifferential and differential Landau-Lifshitz operators for
the case of a normal magnetized homogeneous film with the thickness
$D$. The dispersion relations of spin waves have the form

\begin{equation}
\omega^{(n)2}(\vec
q)=\Omega^{(n)}(\Omega^{(n)}+\Omega_Mq^2/q^{(n)2}_0), \label{eq33}
\end{equation}

\noindent where $n=1,2,3,\ldots$ is the mode number, $\vec q=(q_x,
q_y)$ is the two-dimensional longitudinal wavevector of spin waves,
$q=|\vec q|$, $\Omega^{(n)}=\gamma(H-4\pi M+4\pi \alpha
Mq^{(n)2}_0)$, $\Omega_M= 4\pi\gamma M$, $q^{(n)}_0=(q^2+
q^{(n)2}_z)^{1/2}$, $q_z^{(n)}$ is the transverse vector. The
magnetostatic potential over thickness $z\in[-D/2,D/2]$ of the
magnetic film is

\begin{equation}
\varphi^{(n,\vec
q)}(x,y,z)=(2\pi)^{-1}f^{(n)-1/2}\exp(iq_xx+iq_yy)\cos[q^{(n)}_zz+\pi(n-1)/2],
\label{eq34}
\end{equation}
\noindent where $f^{(n)}=D/2+q/q^{(n)2}_0$.

For the case of pseudodifferential Landau-Lifshitz operators
(\ref{eq31}), the transverse wavevector $ q^{(n)}_z $ is closely
connected to the longitudinal wavevector $q$ by the relation

\begin{equation}
2\cot q^{(n)}_zD=
\frac{q^{(n)}_z}{q}-\frac{q}{q^{(n)}_z}.\label{eq35}
\end{equation}

For the case of differential Landau-Lifshitz operators (\ref{eq32}),
the transverse wavevector is determined by the exchange boundary
conditions and is given by the equation~\cite{Kal86,Gur96}

\begin{equation}
2\cot q^{(n)}_zD=\frac{q^{(n)}_z}{\xi}-\frac{\xi}{
q^{(n)}_z}.\label{eq36}
\end{equation}

Dispersion relations (\ref{eq33}) of the first spin-wave mode
propagating in the YIG film of the thickness $D=$ 0.5~$\mu$m with
$4\pi M=$ 1750~Oe, $\alpha=$ 3.2$\cdot$10${ }^{-12}$~cm${ }^2$ at
the applied magnetic field $H=$ 3000~Oe are shown in figure
\ref{Fig6} for the transverse wave vector $ q^{(1)}_z $ (\ref{eq35})
and for the transverse wave vector $ q^{(1)}_z $ (\ref{eq36}) with
different pinning parameters $\xi$. One can see that there does not
exist any pinning parameter $\xi$, at which the curve $A$ calculated
on the base of relation (\ref{eq35}) coincides with the curves
calculated on the base of the exchange boundary conditions. Thus, we
conclude that the reduction to differential Landau-Lifshitz
operators and the use of the exchange boundary conditions are
incorrect.

\section{Spin-wave relaxation }

In this section we answer the question: what is the value of
spin-wave relaxation in the model with magnetic dipole and exchange
interactions derived from first principles? The answer depends on
the ratio of the spin-wave energy to intervals between modes of the
spin-wave spectrum and is different for thick and for thin magnetic
films. In thick films the spin-wave energy is greater than energy
gaps between modes and a three-spin-wave process takes place. If the
exchange interaction is isotropic, it cannot induce three-magnon
processes and, therefore, the MDI makes a major contribution to the
relaxation. We consider the spin-wave damping in thick films in the
one-loop approximation. In thin magnetic films (for example, in
nanosized films) the energy of long-wavelength spin waves is less
than energy gaps between modes and three-spin-wave processes are
forbidden. In this case, four-spin-wave processes take place, the
exchange interaction makes a major contribution to the relaxation,
and the spin-wave damping has lower values in comparison with the
damping in thick films. We calculate the spin-wave relaxation for
four-spin-wave processes in thin films for long-wavelength spin
waves in the two-loop approximation.

\subsection{Spin-wave relaxation in thick films}

\begin{figure*}
\begin{center}
\includegraphics*[scale=.65]{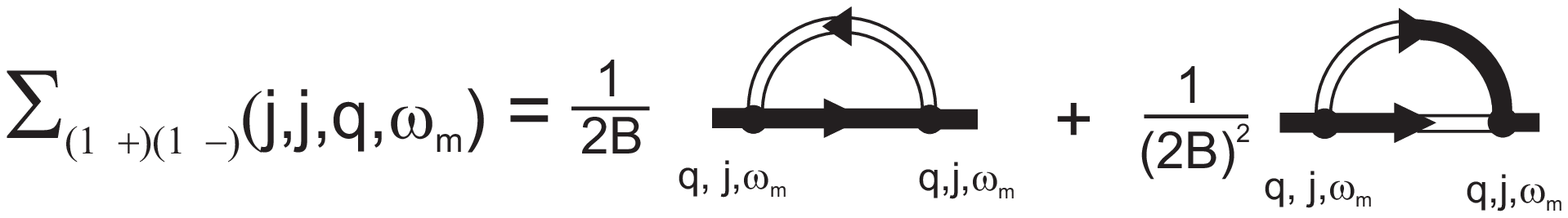}
\end{center}
\caption{Self-energy diagrams in the one-loop approximation at low
temperature. $B$ is determined by relation (\ref{eq8}).
}\label{Fig7}
\end{figure*}

The spin-wave relaxation induced by a three-spin-wave process in
normal magnetized homogeneous ferromagnetic films is considered in
\cite{Lut05,Lut08} in the one-loop approximation for spin waves with
small longitudinal wavevectors at low temperature. The relaxation is
determined by self-energy diagram insertions $\Sigma_{(1{+})(1{-})}$
to the $\bfb P$-matrix given by relation~(\ref{eq12}) (figure
\ref{Fig7}). Damping of the $j$-mode excitation is defined by the
imaginary part of the pole of the effective Green functions $G_{-+}=
P_{(1{-})(1{+})}$ with insertions $\Sigma_{(1{+})(1{-})}$ under the
analytical continuation~(\ref{eq13})

\[\Delta^{(j)}(\vec q)= \frac{\delta\omega^{(j)}(\vec q)}{
\omega^{(j)}(\vec q)}= \left.\frac{2B(p)V_a}{\beta\hbar
\omega^{(j)}(\vec q)} \Ima\Sigma_{(1{+})(1{-})}(j,j, \vec
q,\omega_m) \right|_{i\omega_m\to \omega +
i\varepsilon\signum\omega}\]
\[=\frac{V_a}{2\beta\hbar\omega^{(j)}}\Ima\sum_{n,i,k}\int
F^{(i)} F^{(k)} [\bar P_{(1{-})(1{+})}(i,-\vec q_1,-\omega_n) \bar
P_{(2z)(2z)}(k,\vec q-\vec q_1,\omega_m -\omega_n) \]
\begin{equation}
+\left.\frac1{8B(p)} \bar P_{(1{-})(2z)}(i,\vec q_1,\omega_n) \bar
P_{(2z)(1{+})}(k,\vec q-\vec q_1, \omega_m -\omega_n)] N^2(j,\vec
q;i,\vec q_1;k,\vec q-\vec q_1) \,d^2q_1 \right|_{i\omega_m\to
\omega + i\varepsilon\signum\omega},\label{eq37}
\end{equation}

\noindent where

\[\bar P_{(1{-})(1{+})}\bsp = 2\rho V_a^2(\Omega^{(j)}+2\eta^{(j)}_{{-}{+}}+i\omega_m)\]
\[\bar P_{(1{-})(2z)}\bsp =-2\eta^{(j)}_{{+}{z}}(\Omega^{(j)}+ i\omega_m) \]
\[\bar P_{(2z)(1{+})}\bsp =-2\eta^{(j)}_{z{-}}(\Omega^{(j)}+ i\omega_m) \]
\[\bar P_{(2z)(2z)}\bsp = F^{(j)-1}\beta V_a\tilde
I(q_0^{(j)})- \rho^{-1}\eta^{(j)}_{{z}{z}} (\Omega^{(j)2}+
i\omega_m^2) \]
\[F^{(j)}=(\omega^{(j)2}+\omega^2_m)^{-1},\qquad\rho= \frac{B(p)}{\beta\hbar V_a}, \]
\[\eta^{(j)} _{\mu\nu}=\frac{\Omega_Mq_{\mu} q_{\nu}}{q^{(j)2}_0}\qquad (\mu ,\nu={-},{+},z)\]
\[q_{\pm}=\frac12(q_x\mp iq_y),\]
\[\tilde I(q_0^{(j)})= \tilde I(0)-wq^{(j)2}_0.\]

\noindent is the Fourier transform of the exchange interaction,

\[N(j_1,\vec{q_1};j_2,\vec{q_2};j_3,\vec{q_3})\]
\[= \frac{1}{8\pi V_a}\prod^3_{k=1} \frac{1}{f^{(j_k)1/2}}
\sum_{\sigma_1,\sigma_2,\sigma_3}\frac{\sin\left[\left(\sum^3_{k=1}
\sigma_kq^{(j_k)}_z\right)D/2\right)]}{\sum^3_{k=1}\sigma_kq^{(j_k)}_z}
\exp\left(i\sum^3_{k=1}\sigma_k\pi{(j_k-1)/2}\right) \]

\noindent is the block factor in the representation of the functions
(\ref{eq34}), $f^{(j)}=D/2+q/q_0^{(j)2}$, $\sigma_k=\pm 1$;
$\sum_{\sigma_1,\sigma_2,\sigma_3}$ denotes the summation over all
sets $\{\sigma_1,\sigma_2,\sigma_3\}$. The spin-wave frequency
$\omega^{(j)}$ and the transverse wavevector $ q^{(j)}_z $ are
determined by relations (\ref{eq33}) and (\ref{eq35}), respectively.
The damping $\Delta^{(j)}$ increases directly proportionally to the
temperature.

\begin{figure*}
\begin{center}
\includegraphics*[scale=.5]{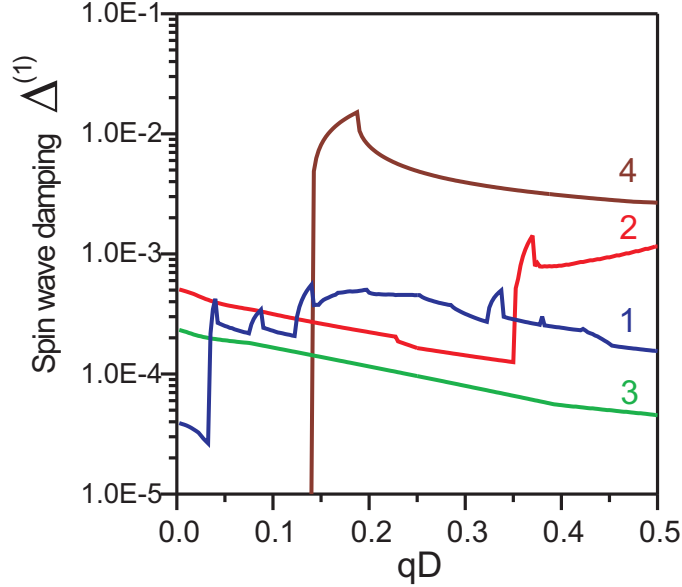}
\end{center}
\caption{Spin wave damping $\Delta^{(1)}=
{\delta\omega^{(1)}}/{\omega^{(1)}}$ of the first mode in normal
magnetized YIG film with the magnetization $4\pi M=$ 1750~Oe and the
exchange interaction constant $\alpha=$ 3.2$\cdot$10${ }^{-12}$~cm${
}^2$ at $H=$ 3000~Oe, $T=$ 300~K at different film thickness $D$.
(1) $D =$ 500~nm, (2) 300~nm, (3) 200~nm, (4) 120~nm. }\label{Fig8}
\end{figure*}

\begin{figure*}
\begin{center}
\includegraphics*[scale=.45]{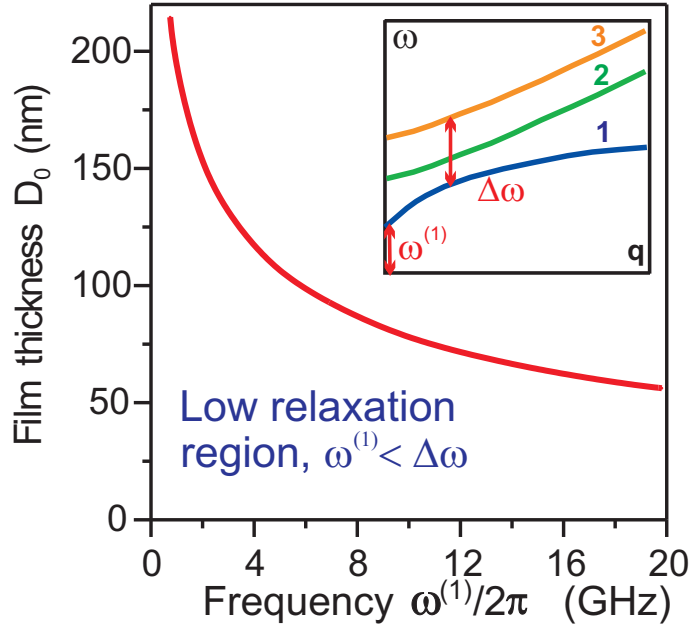}
\end{center}
\caption{Film thickness $D_0$ of YIG film versus the excitation
frequency $\omega^{(1)}(\vec q)/2\pi$ of the first mode at the
wavevector $\vec q\to 0$. Low relaxation region of the first
spin-wave mode exists for YIG films with the thickness
$D<D_0(\omega^{(1)}/2\pi)$. }\label{Fig9}
\end{figure*}

Relation (\ref{eq37}) describes relaxation of the spin-wave $j$-mode
caused by inelastic scattering on thermal excited spin wave modes.
Relaxation occurs through the confluence of the $j$-mode with the
$k$-mode to form the $i$-mode. From the explicit form of the block
factor $N$ in relation (\ref{eq37}) it follows that the confluence
processes take place when the sum of mode numbers $j+i+k$ is equal
to an odd number. The confluence processes are induced by the MDI
and are accompanied by transitions between thermal excited $i$- and
$k$-modes. Transitions take place when equation

\begin{equation}
\omega^{(j)}(\vec q) = \omega^{(i)}({\vec q}{\,}^{(s)})-
\omega^{(k)}(\vec q- {\vec q}{\,}^{(s)}) \label{eq38}
\end{equation}

\noindent has at least one solution ${\vec q}{\,}^{(s)}$ for the
given ${\vec q}$, $i$, $j$,$k$. Existence of solutions ${\vec
q}{\,}^{(s)}$ of equation (\ref{eq38}) depends on the thickness of
the magnetic film. With decreasing film thickness $D$, the density
of dispersion curves of modes on the plane $(\omega ,q)$ decreases
and the frequency of the spacings between curves increase. The least
frequency spacing occurs between the first ($i=1$) and the third
($k=3$) modes. Figure~\ref{Fig8} shows the damping $\Delta^{(1)}$ of
the first spin wave mode versus the longitudinal wave vector $q$
normalized by the film thickness $D$ at different film thicknesses.
Calculations have been done for a YIG film with the magnetization
$4\pi M=$ 1750~Oe and the exchange interaction constant $\alpha=$
3.2$\cdot$10${ }^{-12}$~cm${ }^2$ at $H=$ 3000~Oe and $T=$ 300~K.
One can see that for the YIG film with the thickness $D =$ 120~nm in
the region $qD<0.14$ the damping $\Delta^{(1)}$ is equal to 0 due to
the absence of transitions between modes. Thus, in thin magnetic
films a low spin-wave relaxation region takes place. For the given
$j$-mode this region appears, when the excitation frequency
$\omega^{(j)}(\vec q)$ is less than the difference
$\omega^{(3)}({\vec q}{\,}^{(s)})- \omega^{(1)}(\vec q- {\vec
q}{\,}^{(s)})$ at any values of the wavevector ${\vec q}{\,}^{(s)}$.
For the first mode $\omega^{(1)}$ in the YIG film the low spin-wave
relaxation region is shown in figure~\ref{Fig9} at $q\to 0$. If the
film thickness $D<D_0(\omega^{(1)}/2\pi)$, then $\omega^{(1)}(0) <
\omega^{(3)}({\vec q}{\,}^{(s)})- \omega^{(1)}({\vec q}{\,}^{(s)})$
and the first mode has low values of the spin wave damping
$\Delta^{(1)}$.

\subsection{Relaxation in thin magnetic films}

\begin{figure*}
\begin{center}
\includegraphics*[scale=.65]{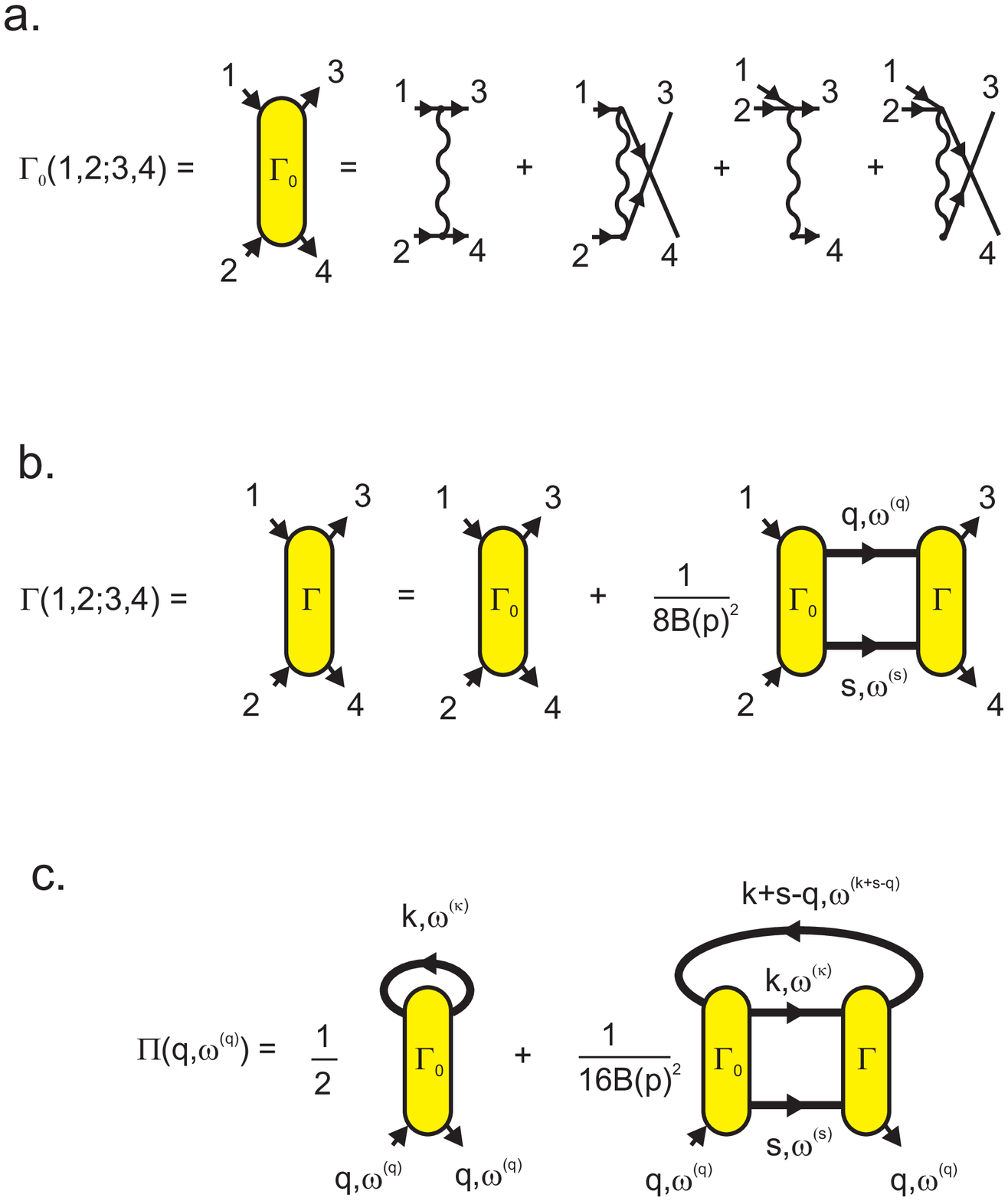}
\end{center}
\caption{(a) Bare $\Gamma_0$-vertex. (b) Ladder approximation. (c)
Self-energy diagram insertion. }\label{Fig10}
\end{figure*}

What is the value of spin wave damping in the low relaxation region
in thin magnetic films? We consider four-spin-wave processes in the
normal magnetized monolayer of the quadratic lattice with the
lattice constant $a$ at small longitudinal wavevector values $\vec
q= (q_x,q_y)$ at low temperature. As isotropy of the exchange
interaction can not forbid four-spin-wave processes and the value of
the exchange interaction much greater than the MDI, only the
exchange interaction will be taken into account in diagrams. We
suppose that the exchange interaction acts between neighboring spins
and is equal to $I_0$. In order to calculate self-energy diagram
insertions to the effective Green functions in the two-loop
approximation, we use the ladder expansion (figure~\ref{Fig10}). At
small values of wavevectors the bare $\Gamma_0$-vertex
(figure~\ref{Fig10}a) is

\[\Gamma_0(1,2;3,4)\equiv\Gamma_0(\vec k,\vec s+\vec q-\vec
k;\vec q,\vec s)\]

\[=\beta[\tilde I(\vec k-\vec q)+\tilde I(\vec k-\vec s)-\tilde I(\vec
s)-\tilde I(\vec q)]= 2\beta I_0a^2(\vec q,\vec s), \]

\noindent where $1,2;3,4$ is the abridged notation of 2-dimensional
wavevectors, which are variables of $\Gamma_0$-vertex; $|\vec
k|,|\vec q|,|\vec s|\ll a^{-1}$;

\[\tilde{I}(\vec q)=\sum_{\vec 1_{xy}-\vec 1'_{xy}}I(\vec 1_{xy}-\vec
1'_{xy})\exp[-i\vec q(\vec 1_{xy}-\vec 1'_{xy})]=
2I_0[\cos(q_xa)+\cos(q_ya)]. \]

\noindent The $\Gamma$-vertex in the ladder approximation
(figure~\ref{Fig10}b) is determined by the relationship

\[\Gamma(1,2;3,4)\equiv\Gamma(\vec k,\omega_1,\vec s+\vec q-\vec
k,\omega_3+\omega_4-\omega_1;\vec q,\omega_3,\vec s,\omega_4)=
\Gamma_0(\vec k,\vec s+\vec q-\vec k;\vec q,\vec s)\]

\[+\frac{1}{8B^2(p)S_b}\sum_{\omega^{(q)}_m}\int \Gamma_0(\vec k,\vec
s+\vec q-\vec k;\vec {q'},\vec s+\vec q-\vec {q'}) G_{{-}{+}}(\vec
{q'},\omega^{(q)}_m) G_{{-}{+}}(\vec s+\vec q-\vec
{q'},\omega_3+\omega_4-\omega^{(q)}_m) \]
\[\times\Gamma(\vec {q'},\omega^{(q)}_m,\vec s+\vec q-\vec {q'},\omega_3+\omega_4
-\omega^{(q)}_m;\vec q,\omega_3,\vec s,\omega_4)\,d^2q',\]

\noindent where

\[G_{{-}{+}}(\vec q,\omega_m)= \frac{2B(p)}{\beta\hbar(\omega(\vec q)-i\omega_m)} \]

\noindent is the effective Green function determined by the ${\bfb
P}$-matrix (\ref{eq12}), $\omega(\vec q)$ is the frequency of spin
excitations in monolayer (\ref{eq19}), $S_b$ is the volume of the
2-dimensional first Brillouin zone. The coefficient $1/8B^2(p)$ is
due to the fact that the substitution of the bare Green function to
effective ones in diagrams are performed inside blocks. The
self-energy diagram insertion (figure~\ref{Fig10}c) is given by

\[\Pi(\vec q,\omega^{(q)}_m)= \frac{1}{2S_b}\sum_{\omega^{(k)}_n}\int \Gamma_0(\vec q,
\vec k;\vec q,\vec k) G_{{-}{+}}(\vec k,\omega^{(k)}_n)\,d^2k \]

\[+\frac{1}{16B^2(p)S_b^2}\sum_{\omega^{(k)}_n,\omega^{(s)}_l}\int\int \Gamma_0(\vec q,
\vec s+\vec k-\vec q;\vec s,\vec k) G_{{-}{+}}(-\vec s-\vec k+\vec
q,-\omega^{(k)}_n-\omega^{(s)}_l+\omega^{(q)}_m) \]

\begin{equation}\times G_{{-}{+}}(\vec k,\omega^{(k)}_n) G_{{-}{+}}(\vec s,\omega^{(s)}_l)
\Gamma(\vec s,\omega^{(s)}_l,\vec k,\omega^{(k)}_n;\vec
q,\omega^{(q)}_m,\vec s+\vec k-\vec
q,\omega^{(k)}_n+\omega^{(s)}_l-\omega^{(q)}_m)\,d^2k\,d^2s.
\label{eq39}
\end{equation}

\noindent The damping of spin wave excitations is expressed by the
imaginary part of the self-energy $\Pi(\vec q,\omega^{(q)}_m)$

\begin{equation}
\Delta(\vec q)= \frac{\delta\omega(\vec q)}{ \omega(\vec q)}=
\left.\frac{\Ima\Pi(\vec q,\omega^{(q)}_m)}{\beta
\omega}\right|_{i\omega_m\to \omega + i\varepsilon\signum\omega}.
\label{eq40}
\end{equation}

\noindent Taking into account the self-energy $\Pi(\vec
q,\omega^{(q)}_m)$ in the Born approximation, namely, substituting
$\Gamma\rightarrow\Gamma_0$ in relation (\ref{eq39}), integrating
over $\vec k$, $\vec s$ and summing over the frequency variables
$\omega^{(k)}_n$ and $\omega^{(s)}_l$ in equation (\ref{eq40}), at
$\hbar\omega (\vec q)<kT$ we obtain

\[\Delta(\vec q)= \frac{C(qa)^2(kT)^2}{16\pi B^2(p)I_0\varepsilon^{(0)}}, \]

\noindent where $C=1.12$, $k$ is the Boltzmann constant,
$\varepsilon^{(0)}=\hbar\gamma (H+H^{(m)})$ is the Zeeman energy. In
order to evaluate the damping of spin waves, we calculate
$\Delta(\vec q)$ for spin waves with the wavelength $\lambda=$
5~$\mu$m propagating in the monolayer film with the lattice constant
$a=$ 0.4~nm and with the exchange interaction between neighboring
spins $I_0=$ 0.085~eV, $B(p)=1/2$ at $T=$ 300~K. Then, taking into
account that $q=2\pi/\lambda$, for $\varepsilon^{(0)}/h=$ 10~GHz we
obtain $\Delta(\vec q)=$ 4.28$\cdot 10^{-6}$. Thus, one can see that
the damping of spin waves of small wavevectors is low.

\section{Spin-wave devices on the base of nanosized magnetic films}

According to the above-mentioned, spin excitations in nanosized
films have low damping. This property can be used in spin-wave
devices. We consider spin-wave filters on the base of nanosized
magnetic films and field-effect transistors with magnetic films
under the gate contact.

\subsection{Spin-wave filters}

\begin{figure*}
\begin{center}
\includegraphics*[scale=.65]{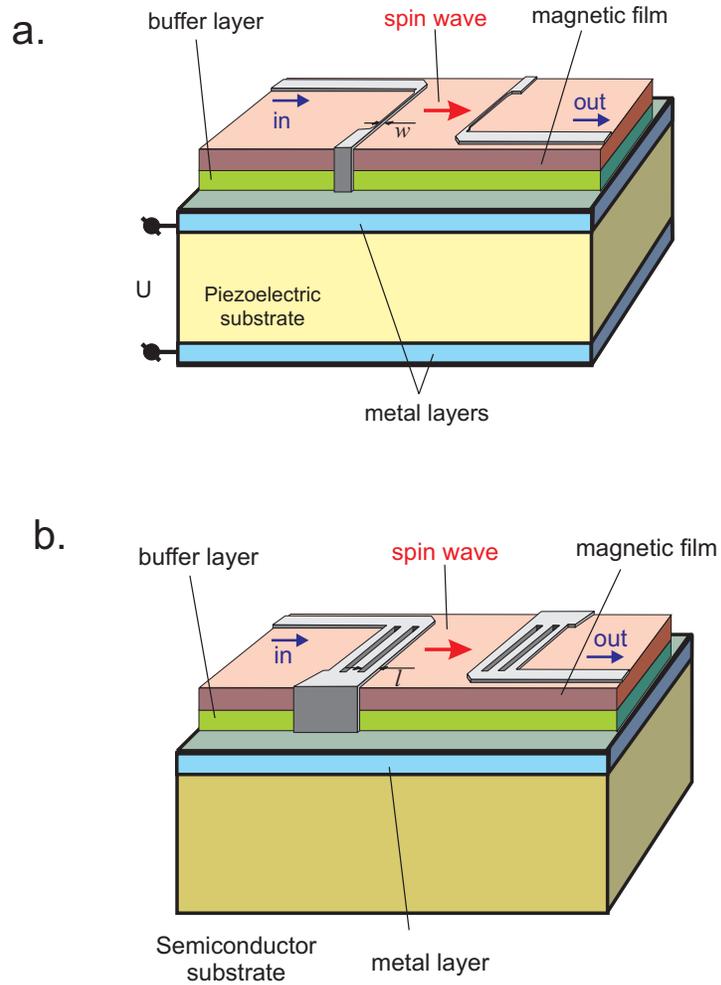}
\end{center}
\caption{(a). Spin-wave filter on the base of nanosized magnetic
film on a piezoelectric substrate. (b). Spin-wave filter with
nanosized magnetic film on a semiconductor substrate. }\label{Fig11}
\end{figure*}

Using thin magnetic films, we can construct spin-wave devices of
small sizes and can integrate them to semiconductor chips. Figure
\ref{Fig11}a presents tunable spin-wave filter on a piezoelectric
substrate. Microwave frequency current flowing in the microstrip
antenna of the width $w$ generates spin waves in the magnetic film.
Wavevector of spin waves is determined by the antenna width and is
in the range [0, $2\pi /w$]. When spin waves come up to the second
antenna, the magnetic field of spin waves induces a current of the
same frequency in this antenna. The waveguide impedance of the
filter depends on the antenna width, the width of microstrips, the
dielectric constant of the film, and the thickness of the film
between microstrips and the upper metal contact on the piezoelectric
substrate. In order to remove a lattice mismatch and to reach
desirable impedance, buffer layers between the magnetic film and the
metal contact can be used. Tunability of the filter is provided by
lattice variations of the piezoelectric substrate. The applied
voltage $U$ varies the lattice constant of the substrate and, as a
result of this variation, varies the lattice constant of the
magnetic film. Compression and expansion of the lattice lead to the
stress anisotropy $H^{(a)}$ in the magnetic film~\cite{Krup}. In
this case, we must substitute $H\rightarrow H+H^{(a)}$ in the spin
propagators (\ref{eq7}), in the Green functions (\ref{eq11}), in the
frequency $\Omega(\vec q)$ in (\ref{eq19}), (\ref{eq20}), and in the
Landau-Lifshitz equations (\ref{eq28}), (\ref{eq29}). The frequency
$\omega$ of spin waves is varied.

Narrow-band filters can be constructed on the base of periodic
antennae (Figure \ref{Fig11}b). These antenna structure generate
spin waves with the wavevector $q=2\pi/l$, where $l$ is the period
of the generating antenna. Filters with periodic antenna structures
have more selectivity in comparison with filters with single
antennae. The bandwidth of the filter is given by

\[\Delta\omega=\left.\frac{\partial\omega(q)}{\partial
q}\right|_{q_0}\cdot\frac{2\pi}{lN_1N_2}\]

\noindent where $\omega(q)$ is dispersion relation (\ref{eq19}),
(\ref{eq20}), or (\ref{eq33}), $q_0=2\pi/l$, $N_1$, $N_2$ are
numbers of periods of generating and receiving antennae,
respectively. Thin magnetic film used in the above-mentioned filters
must be dielectric and can be garnet, spinel or hexaferrite films.
These films can be produced by the laser deposition method or by
ion-beam sputtering with following annealing. In \cite{Man09} band
pass spin-wave filters at $5-7$ GHz have been fabricated on the base
of submicron thick YIG films produced by the laser deposition on
Gd${ }_3$Ga${ }_5$O${ }_{12}$ substrates. In the case, when
spin-wave filters are integrated on semiconductor chips, the
semiconductor must endure the annealing procedure without any
changes for the worse in the semiconductor structure. For this
purpose, Si and GaN can be used.

\subsection{Field-effect transistors with nanosized magnetic films}

\begin{figure*}
\begin{center}
\includegraphics*[scale=.55]{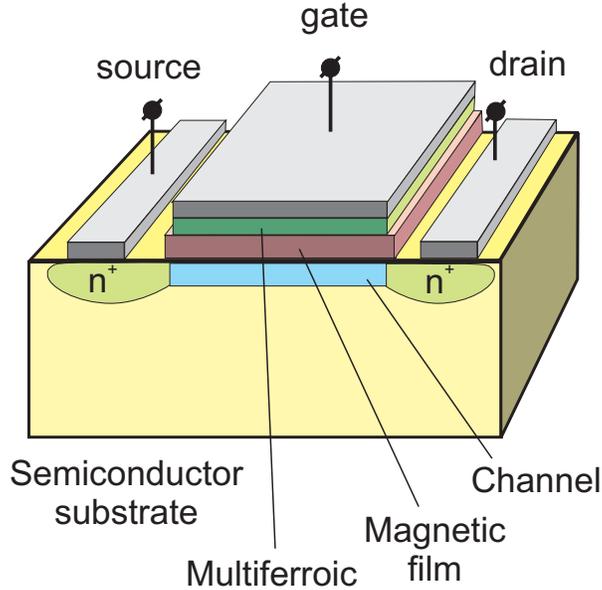}
\end{center}
\caption{Field-effect transistor with nanosized magnetic film under
the gate contact. }\label{Fig12}
\end{figure*}

As we can see from relations (\ref{eq20}), (\ref{eq21}) and from
figures \ref{Fig4}, \ref{Fig5}, spin-wave resonance peaks in thin
magnetic films have high frequencies and low numbers. For the
exchange interaction between layers $I_d=$ 0.05-0.1~eV (garnet
films, lithium ferrospinel and hexaferrite films \cite{Krup,Har})
spin-wave resonance frequencies are in Gigahertz and Terahertz
frequency bands. Distance between spin-wave resonance peaks is
greater in thin films than in thick ones. Since in thin films
spin-wave resonances (\ref{eq21}) at the given high frequency have
lower numbers $n$ and have lower numbers of half-periods of waves on
film thickness in comparison with resonances at the same frequency
in thick magnetic films, high-frequency resonances in thin films can
be exited more easily. This can be used in field-effect transistor
(FET) structures. In figure \ref{Fig12} thin dielectric magnetic
film is placed under the gate electrode. Applied voltage of the
microwave frequency on the gate generates electromagnetic field,
which induces spin-wave resonances. In order to increase the
magnetic component of the electromagnetic field and to enhance
amplitude of resonances, multiferroic layer can be placed between
the gate and the magnetic film. The magnetic field of the induced
spin-wave resonances interacts with spins of electrons propagating
in the channel and modulates the current of the transistor.
Consequently, existence of spin-wave resonances in the gate circuit
can be regarded as a filter. Choosing the certain spin-wave
resonance peak, we can construct FET structure operating in a
desirable range of Gigahertz and Terahertz frequency bands.

\section{Conclusions}

The results of the investigations can be summarized as follows.

\noindent (1) We have studied spin excitations in nanosized magnetic
films in the Heisenberg model with magnetic dipole and exchange
interactions by the spin operator diagram technique. Dispersion
relations of spin waves in two-dimensional magnetic monolayers and
in two-layer magnetic films and the spin-wave resonance spectrum in
$N$-layer structures are found.

\noindent (2) Generalized Landau-Lifshitz equations for thick
magnetic films which are derived from first principles, have the
integral (pseudodifferential) form, but not differential one with
respect to spatial variables. Spin excitations are determined by
simultaneous solution of the Landau-Lifshitz equations and the
equation for the magnetostatic potential. The use of exchange
boundary conditions for solvability of the Landau-Lifshitz equations
is incorrect.

\noindent (3) The magnetic dipole interaction makes a major
contribution to the relaxation of long-wavelength spin waves in
thick magnetic films. The spin-wave damping is determined by
diagrams in the one-loop approximation, which correspond to
three-spin-wave processes. The three-spin-wave processes are
accompanied by transitions between thermal excited spin-wave modes.
The damping increases directly proportionally to the temperature.

\noindent (4) Thin films have a region of low relaxation of
long-wavelength spin waves. In thin magnetic films the energy of
these waves is less than energy gaps between spin-wave modes,
therefore, three-spin-wave processes are forbidden, four-spin-wave
processes take place and, as a result of this, the exchange
interaction makes a major contribution to the relaxation. It is
found that the damping of spin waves propagating in a magnetic
monolayer has the form of the quadratic dependence on the
temperature and is very low for spin waves with small wavevectors.

\noindent (5) Nanosized magnetic films can be used in spin-wave
devices. Low damping of long-wavelength spin waves gives us
opportunity to construct tunable narrow-band spin-wave filters with
high quality at the microwave frequency range. Spin-wave resonances
of nanosized magnetic films under gate electrodes in field-effect
transistor (FET) structures can be used to construct FET structures
operating in Gigahertz and Terahertz frequency bands.

\section*{Acknowledgment}
This work was supported by the Russian Foundation for Basic
Research, grant 10-02-00516, and by the Ministry of Education and
Science of the Russian Federation, project 2011-1.3-513-067-006.


\end{document}